\documentclass[a4paper,12pt]{article}

\pdfoutput=1
\usepackage{jheppub}
\usepackage[T1]{fontenc}

\usepackage{graphicx}
\usepackage{cancel}
\usepackage{amssymb}
\usepackage{textcomp}
\usepackage{amsmath}
\usepackage{bm}
\usepackage{times}
\usepackage{epsfig}
\usepackage{epstopdf}
\usepackage{color}
\usepackage{multirow}

\def\lsim{\mathrel{\raise.3ex\hbox{$<$\kern-.75em\lower1ex\hbox{$\sim$}}}}
\def\gsim{\mathrel{\raise.3ex\hbox{$>$\kern-.75em\lower1ex\hbox{$\sim$}}}}

\def\beq{\begin{equation}}
\def\eeq{\end{equation}}
\def\be{\begin{equation}}
\def\ee{\end{equation}}
\def\bea{\begin{eqnarray}}
\def\eea{\end{eqnarray}}

\def\ev{\,{\rm eV}}
\def\to{\rightarrow}

\newcommand{\minigraph}[5][0.25in]{\begin{minipage}{#2}\begin{center}\includegraphics[width=#2]{#5}\\\vspace{#3}\hspace{#1}{\footnotesize #4}\end{center}\end{minipage}}

\title{Type II Seesaw and tau lepton at the HL-LHC, HE-LHC and FCC-hh}

\author{Tong Li}

\emailAdd{tong.li@monash.edu}
\affiliation{
ARC Centre of Excellence for Particle Physics at the Tera-scale, School of Physics and Astronomy, Monash University, Melbourne, Victoria 3800, Australia}

\abstract{
The tau lepton plays important role in distinguishing neutrino mass patterns and determining the chirality nature in heavy scalar mediated neutrino mass models, in the light of the  neutrino oscillation experiments and its polarization measurement. We investigate the lepton flavor signatures with tau lepton at LHC upgrades, i.e. HL-LHC, HE-LHC and FCC-hh, through leptonic processes from doubly charged Higgs in the Type II Seesaw. We find that for the channel with one tau lepton in final states, the accessible doubly charged Higgs mass at HL-LHC can reach 655 GeV and 695 GeV for the neutrino mass patterns of normal hierarchy (NH) and inverted hierarchy (IH) respectively, with the luminosity of 3000 fb$^{-1}$. Higher masses, 975-1930 GeV for NH and 1035-2070 GeV for IH, can be achieved at HE-LHC and FCC-hh.
}

\begin{document}

\maketitle
\flushbottom
\newpage

\section{Introduction}

It is well-known that, in the context of the Standard Model (SM), the small but non-zero Majorana neutrino masses can be realized at leading order through a dimension-5 operator~\cite{Weinberg}
\begin{eqnarray}
{\kappa\over \Lambda}l_Ll_LHH
\label{weinberg}
\end{eqnarray}
where $l_L$ and $H$ stand for the SM lepton doublet and the Higgs doublet, respectively.
Among the only three ultraviolet completions of Eq.~(\ref{weinberg}) at tree level, known as the Type I, Type II and Type III Seesaw mechanisms, the Type II Seesaw introduces an SU$(2)$ Higgs triplet $\Delta$ and the consequent doubly charged Higgs ($H^{\pm\pm}$) decays with the violation of lepton number after the electroweak symmetry breaking (EWSB)~\cite{Konetschny:1977bn, Cheng:1980qt,Lazarides:1980nt,Schechter:1980gr,Mohapatra:1980yp}. If the lepton number violation happens at TeV scale, the Seesaw mechanism holds in terms of a relevant small coupling of the dimension-5 operator~\cite{deGouvea:2006gz,deGouvea:2007hks} and the new Higgs states can be experimentally accessible at the Large Hadron Collider (LHC)~\cite{LNVreview}. Up to now, no significant evidence of doubly charged Higgs was observed at the LHC and thus the lower limit on doubly charged Higgs mass ($M_{H^{\pm\pm}}$) emerged. 
Assuming the branching fraction of $H^{\pm\pm}$ decay into same sign dilepton to be ${\rm BR}(H^{\pm\pm}\to \ell^\pm\ell^\pm)=100\%$ and $10\%$, the most stringent lower mass limit by ATLAS is $770-870$ GeV and 450 GeV respectively~\cite{Aaboud:2017qph}, with 36.1 fb$^{-1}$ of integrated luminosity at 13 TeV LHC.

The relevant signal regions analyzed by ATLAS were defined by the combination of electron and/or muon final states. In other words, ATLAS only considered the decays of $H^{\pm\pm}$ such as $H^{\pm\pm}\to e^\pm e^\pm$, $\mu^\pm \mu^\pm$ or $e^\pm \mu^\pm$. It was emphasized that, however, governed by the constraints from neutrino oscillation experiments, the channels with tau lepton in doubly charged Higgs decays play an important role in distinguishing neutrino mass hierarchies and determining mixing parameters~\cite{Perez:2008ha,LNVreview}.
By contrast to ATLAS, CMS recently performed doubly charged Higgs search using events with tau lepton at the center-of-mass energy of 13 TeV with 12.9 fb$^{-1}$ luminosity~\cite{CMS:2017pet}. The lower bound on the $H^{\pm\pm}$ mass was reported to be about 481-537 (396) GeV assuming ${\rm BR}(H^{\pm\pm}\to \tau^\pm \ell^\pm (\tau^\pm\tau^\pm))=100\%$. In the CMS search and relevant studies~\cite{delAguila:2013mia}, hadronic tau leptons were reconstructed without considering their polarization as chirality does not play any role in their analyses. The tau leptons before decay are considered to be tagged by a geometrical method.

Actually, the measurement of chirality in doubly charged Higgs search with tau lepton is important as it can help to discriminate different heavy scalar mediated neutrino mass mechanisms in which the doubly charged Higgs can couple to either left-handed or right-handed leptons. For example, $H^{--}$ from an SU$(2)_L$ triplet field in the Type II Seesaw model only couples to left-handed charged leptons, while the one from an SU$(2)_L$ singlet field in the Zee-Babu model~\cite{Zee:1985id,Babu:1988ki} only interacts with right-handed charged leptons.
Without the chirality measurement in doubly charged Higgs search, it is impossible to discriminate these models except for their different production cross sections~\cite{delAguila:2013mia,Aaboud:2017qph}.
Fortunately, it was pointed out that the chiral properties of the doubly charged Higgs Yukawa interactions can be determined by looking at the distributions of tau leptons' decay products~\cite{Sugiyama:2012yw}. Hence, it is necessary to investigate the search for doubly charged Higgs at colliders taking into account of both the neutrino oscillation constraints and the tau polarization.

Recently, several precise reactor measurements lead to more accurate lepton flavor predictions about the Seesaw models. For instance, Double Chooz~\cite{Abe:2011fz}, RENO~\cite{Ahn:2012nd} and in particular Daya Bay~\cite{An:2012eh}, have reported non-zero measurements of $\theta_{13}$ by looking for the disappearance of anti-electron neutrino. T2K and NOvA reported on indications of a non-zero leptonic CP phase~\cite{Abe:2013hdq,Adamson:2016tbq,Abe:2017uxa}. These experiments provide us up-to-date neutrino oscillation results
to investigate the impact on neutrino Seesaw models and consequently examine the lepton flavor signatures to be searched at colliders.
Moreover, tau decay packages like Tauola~\cite{Jadach:1990mz,Jezabek:1991qp,Jadach:1993hs} and more recent Taudecay~\cite{Hagiwara:2012vz} were developed to handle tau lepton decay. Thus, a detailed assessment of the search sensitivity is timely in order to seriously
consider the channels with tau lepton from the experimental point of view, at the LHC upgrades such as the High-Luminosity LHC (HL-LHC) and the Higher-Energy LHC (HE-LHC) and the future 100 TeV pp circular collider (FCC-hh).

This paper is organized as follows. In Sec.~\ref{sec:tauLHC}, we first outline the Type II Seesaw model, discuss the constraints from neutrino oscillation experiments and show different doubly charged Higgs decay patterns.
Then we simulate the production of doubly charged Higgs and its lepton number violating decays with tau lepton(s) in the final states at the LHC.
The results of projected search for doubly charged Higgs search using tau lepton at HL-LHC, HE-LHC, and FCC-hh are given in Sec.~\ref{sec:results}.
Finally, in Sec.~\ref{sec:Concl} we summarize our conclusions.

\section{Type II Seesaw with tau lepton at the LHC}
\label{sec:tauLHC}

\subsection{Type II Seesaw model}

We first review the lepton flavor physics in the Type II Seesaw mechanism.
In Type II Seesaw an SU$(2)_L$ scalar triplet $\Delta\sim (1,3,1)$ which can be decomposed as
\begin{eqnarray}
\Delta= \left(
  \begin{array}{cc}
    \delta^+/\sqrt{2} & \delta^{++} \\
    \delta^0 & -\delta^+/\sqrt{2} \\
  \end{array}
\right)
\end{eqnarray}
interacts with SM lepton doublet $l_L$ through a Yukawa coupling $Y_\nu$
\begin{eqnarray}
Y_\nu \ l^T_L \ C\ i\sigma_2 \ \Delta \ l_L+h.c. \ \ ,
\end{eqnarray}
and it also couples with the SM Higgs doublet $H$ via the mixing term
\begin{eqnarray}
\mu H^T \ i\sigma_2 \ \Delta^\dagger H+h.c. \ .
\end{eqnarray}
The neutrino mass is then given by
\begin{eqnarray}
M_\nu=\sqrt{2}Y_\nu v_\Delta, \ \ \ v_\Delta={\mu v_0^2\over \sqrt{2} M_\Delta^2}  \ ,
\end{eqnarray}
with $M_\Delta$ being the mass of the heavy triplet Higgs, and
$v_0$ and $v_\Delta$ are the vevs of the neutral component of the Higgs doublet and triplet respectively satisfying
$v_0^2+v_\Delta^2\approx (246 \ {\rm GeV})^2$. As a result, the lepton number is broken
by $\Delta$ spontaneously. After the electroweak
symmetry breaking, there are seven physical Higgses, including the
singly charged Higgs $H^\pm\approx \delta^\pm$ and doubly charged Higgs $H^{\pm\pm}=\delta^{\pm\pm}$ with $M_{H^\pm}=M_{H^{\pm\pm}}=M_\Delta$.
The Yukawa interactions of the doubly charged Higgs are
\begin{align}
\ell_L^T \ C \ Y^{++}_\nu \ H^{++} \ \ell_L,  \qquad &Y^{++}_\nu = {M_\nu\over \sqrt{2} v_\Delta} = U_{PMNS}^* \ \frac{m_{\nu}^{diag}}{\sqrt{2} \ v_{\Delta}} \ U_{PMNS}^{\dagger},
\label{Y++}
\end{align}
with $U_{PMNS}$ being the Pontecorvo-Maki-Nakagawa-Sakata (PMNS) neutrino mixing matrix and the partial width of doubly charged Higgs decay into same-sign leptons is thus given by
\begin{eqnarray}
\Gamma(H^{++}\to \ell^+_i \ell^+_j)={1\over 4\pi(1+\delta_{ij})}|(Y^{++}_{\nu})_{ij}|^2 M_{H^{++}}.
\label{width}
\end{eqnarray}
Note that below $v_{\Delta}\approx 10^{-4}$ GeV,
the decays of doubly charged Higgs $H^{++}$ are dominated by the above lepton number violating channels~\cite{Perez:2008ha}.

In order to understand the implication of the neutrino experiments,
we then discuss the neutrino mass and mixing parameters in the light of oscillation data.
The neutrino mixing matrix can be parameterized as
\beq U_{PMNS}= \left(
\begin{array}{lll}
 c_{12} c_{13} & c_{13} s_{12} & e^{-\text{i$\delta $}} s_{13}
   \\
 -c_{12} s_{13} s_{23} e^{\text{i$\delta $}}-c_{23} s_{12} &
   c_{12} c_{23}-e^{\text{i$\delta $}} s_{12} s_{13} s_{23} &
   c_{13} s_{23} \\
 s_{12} s_{23}-e^{\text{i$\delta $}} c_{12} c_{23} s_{13} &
   -c_{23} s_{12} s_{13} e^{\text{i$\delta $}}-c_{12} s_{23} &
   c_{13} c_{23}
\label{PMNS}
\end{array}
\right)\times \text{diag} (e^{i \Phi_1/2}, 1, e^{i \Phi_2/2}) \eeq
where $s_{ij}\equiv\sin{\theta_{ij}}$, $c_{ij}\equiv\cos{\theta_{ij}}$, $0 \le
\theta_{ij} \le \pi/2$ and $0 \le \delta, \Phi_i \le 2\pi$ with $\delta$ being the Dirac CP phase and $\Phi_i$ the Majorana phases.
The size of the mass-squared splitting between three
neutrino states is extracted from neutrino oscillation experiments.
The sign of $\Delta m_{31}^2 = m_{3}^2 - m_{1}^2$, however, still remains unknown, which can be either positive, the
Normal Hierarchy (NH), or negative, the Inverted Hierarchy (IH), for the spectrum of the neutrino masses.

Taking into account the reactor data from the antineutrino disappearance experiments mentioned above together with other disappearance and appearance results, the latest global fit of the neutrino masses and mixing parameters, at $3\sigma$ level~\cite{Esteban:2016qun}, are listed here
\bea
6.8 \times 10^{-5} \ev^2 \  < & \Delta m_{21}^2 & < \  8.02 \times 10^{-5} \ev^2, \nonumber \\
2.399 \times 10^{-3} \ev^2 \  < & \Delta m_{31}^2 & < \  2.593 \times 10^{-3} \ev^2 , \nonumber \\
(-2.562 \times 10^{-3} \ev^2 \  < & \Delta m_{32}^2 & < \  -2.369 \times 10^{-3} \ev^2 ), \nonumber \\
                   0.272 \  < & \sin^2{\theta_{12}} & < \  0.346, \nonumber \\
                   0.418 \ (0.435) \  < & \sin^2{\theta_{23}} & <\  0.613 \ (0.616), \nonumber \\
                          0.01981 \ (0.02006)\ < & \sin^2{\theta_{13}} & <\  0.02436 \ (0.02452) , \nonumber \\
                          144^\circ \ (192^\circ)\ < & \delta_{\rm CP} & <\ 374^\circ (354^\circ),
\eea
for NH (IH).
We also adopt the tightest constraint on the sum of neutrino masses
by combining the Planck+WMAP+highL+BAO data~\cite{Ade:2015xua} at $95\%$ confidence level (CL),
\beq
\sum_{i=1}^3 m_i < \ 0.230 \ \ev .
\eeq
Using the above experimental constraints and the decay width formula in Eq.~(\ref{width}) we can obtain the allowed values for
different doubly charged Higgs decay patterns for NH and IH below.

\begin{figure}[h!]
\begin{center}
\minigraph{6cm}{-0.15in}{(a)}{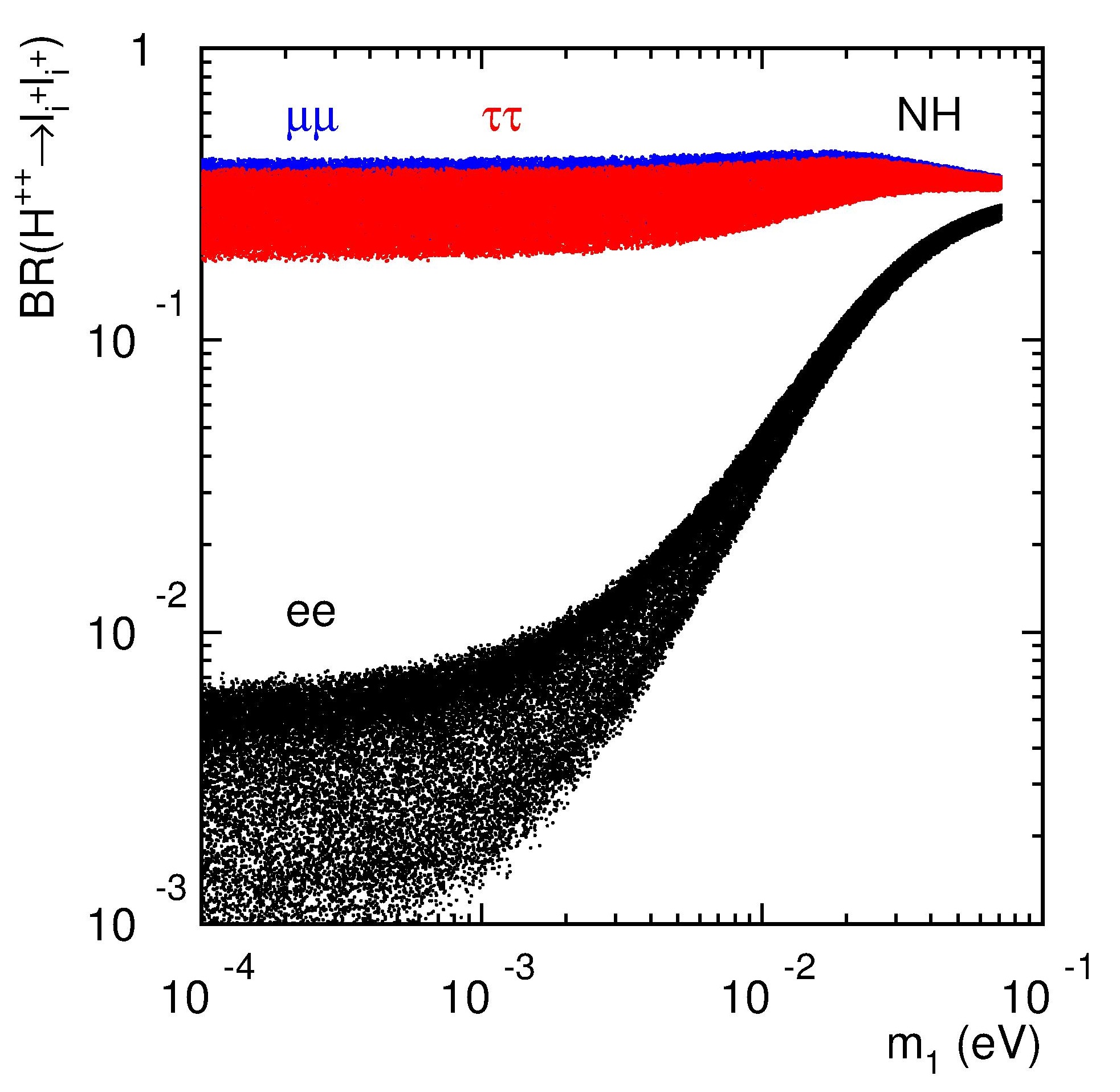}
\minigraph{6cm}{-0.15in}{(b)}{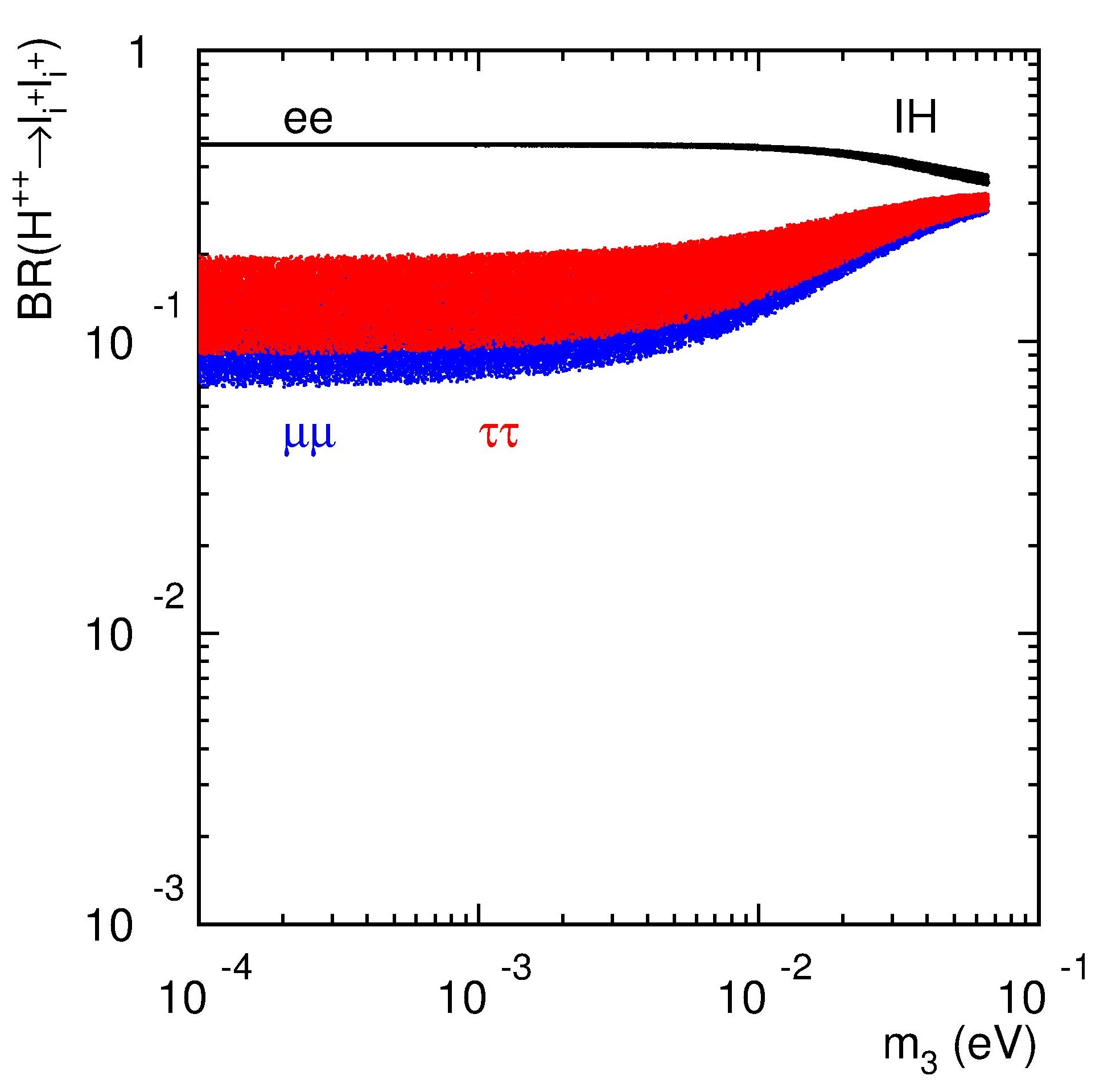}
\minigraph{6cm}{-0.15in}{(c)}{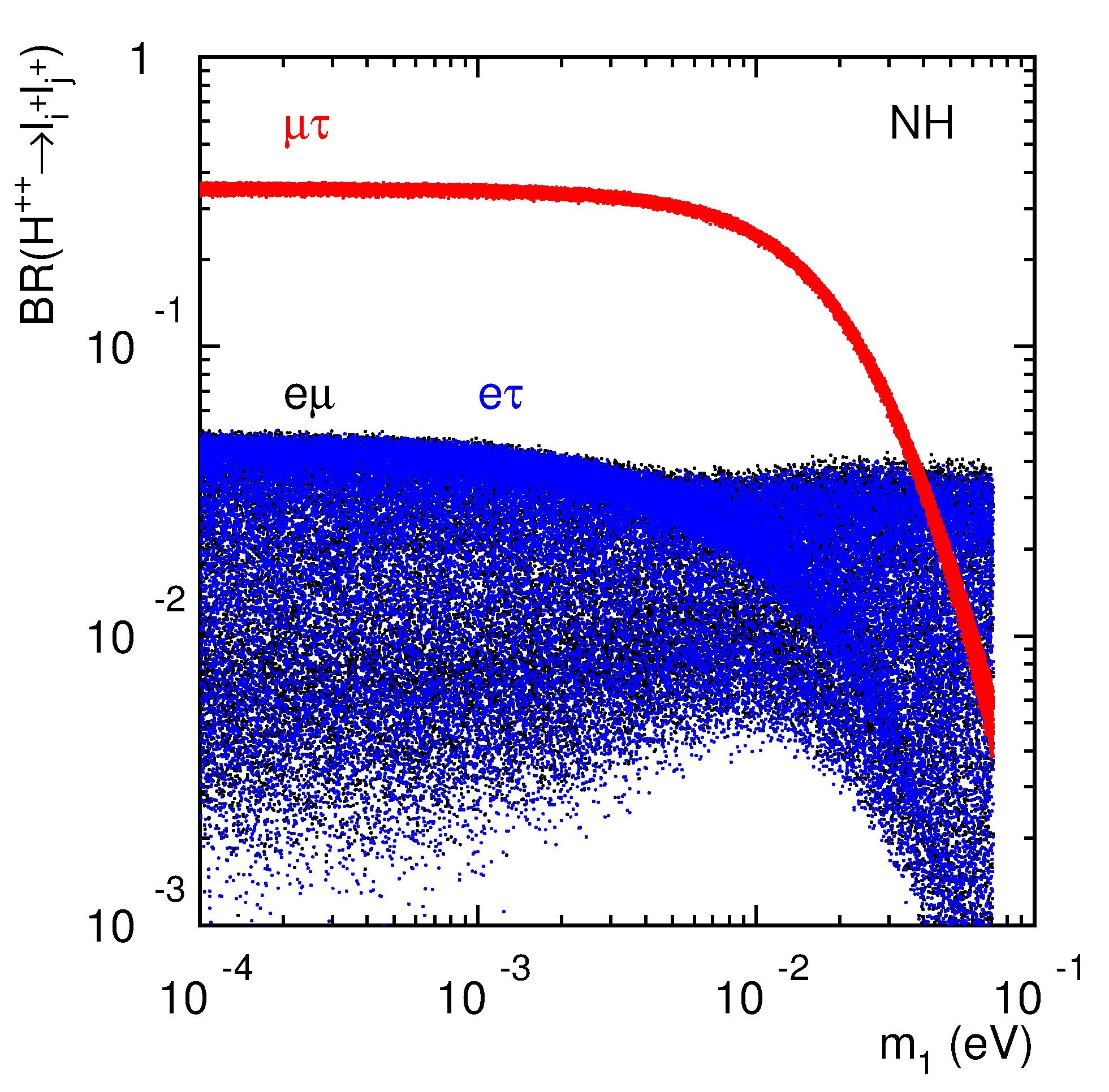}
\minigraph{6cm}{-0.15in}{(d)}{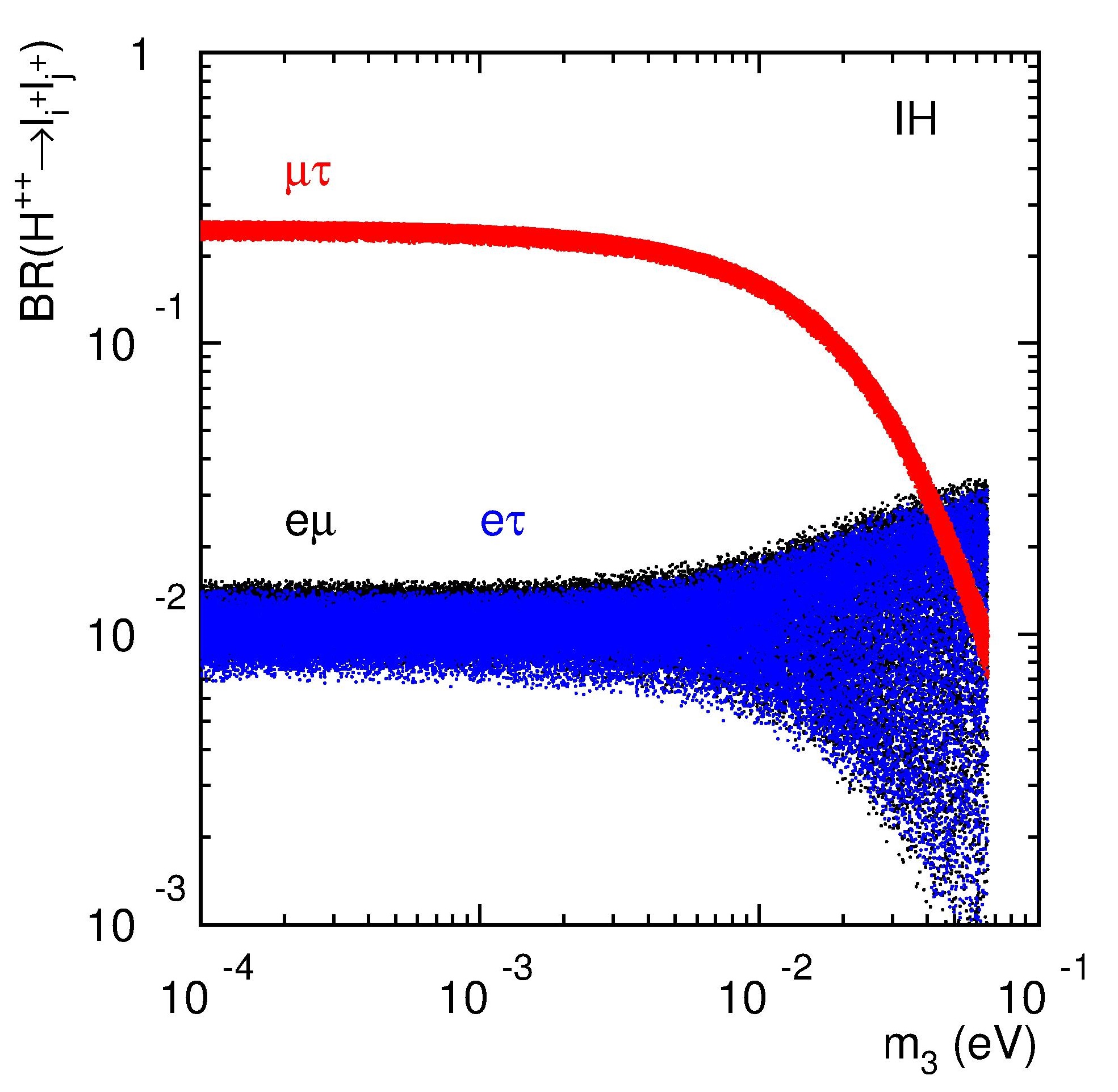}
\end{center}
\caption{Scatter plots for the $H^{++}$ decay branching fractions
 to the flavor-diagonal (a, b) and flavor-off-diagonal (c, d) like-sign dileptons versus the lowest
neutrino mass for NH (a, c) and IH (b, d) with $\Phi_1 = \Phi_2 = 0$.}
\label{scatterBR}
\end{figure}

For simplicity, we ignore the effects of the Majorana phases, i.e. $\Phi_1 = \Phi_2 = 0$ in Eq.~(\ref{PMNS}). In the case of the decays into two identical leptons as shown in Fig.~\ref{scatterBR} (a, b), in the NH case, the branching fraction of
${\rm BR}(H^{++}\to \tau^+\tau^+)$ is comparable to that of $\mu^+\mu^+$ channel and differs from ${\rm BR}(H^{++}\to e^+e^+)$ by two orders of magnitude.
In the neutrino mass pattern of IH, the branching ratios of $e^+ e^+,\mu^+ \mu^+,\tau^+ \tau^+$ channels remain in the same order. For both spectra in the case of the decays with different lepton flavors in the final states, $H^{++}\to \tau^+\mu^+$ is always dominant with at least one order of magnitude larger branching ratio, compared to $e^+\mu^+,e^+\tau^+$ as shown in Fig.~\ref{scatterBR} (c, d). In our numerical calculations below, we take two benchmarks of doubly charged Higgs decay branching ratios for NH and IH, as shown in Table~\ref{BR}. They are consistent with the scatter plots in Fig.~\ref{scatterBR}.

\begin{table}[tb]
\begin{center}
\begin{tabular}{|c|c|c|c|c|c|c|}
\hline
BR  & $ee$ & $e\mu$ & $e\tau$ & $\mu\mu$ & $\mu\tau$ & $\tau\tau$
\\ \hline
NH & 0 & 2.5\% & 2.5\% & 30\%  & 35\% & 30\%
\\ \hline
IH & 50\% & 1\% & 1\% & 12\%  & 24\% & 12\%
\\ \hline
\end{tabular}
\end{center}
\caption{Benchmark decay branching ratios of doubly charged Higgs for NH and IH spectra.}
\label{BR}
\end{table}

\subsection{Testing Type II Seesaw with tau lepton at the LHC}

The most appealing production at hadron colliders for
triplet Higgs bosons is the pair production of doubly charged Higgs
\begin{eqnarray}
&&pp\to Z^\ast/\gamma^\ast \to H^{++} H^{--}
\end{eqnarray}
followed by lepton number violating decays as discussed in last section.
We show the total cross section of $pp\to H^{++} H^{--}$ with collision energy of 14 TeV, 27 TeV and 100 TeV respectively, through $q\bar{q}$ annihilation, and apply an overall next-to-leading (NLO) QCD K-factor of 1.25 below~\cite{Muhlleitner:2003me}.
The main backgrounds come from diboson and $t\bar{t}X$ channels~\cite{Aaboud:2017qph}.
Without loss of generality, we choose
$ZZ$ and $t\bar{t}Z$ events to estimate our background contribution as they produce irreducible background source.
With NLO in QCD their cross sections are $\sigma(ZZ)=14.4\pm 0.1$ pb and $\sigma(t\bar{t}Z)=0.664\pm 0.006$ pb at 14 TeV LHC using MadGraph5\_aMC@NLO~\cite{Alwall:2014hca}. LNV-Scalars\_UFO~\cite{delAguila:2013mia} is interfaced with MadGraph5 to generate signal events. We also use Taudecay\_UFO~\cite{Hagiwara:2012vz} to simulate tau lepton decay carrying polarization information at parton level.
From the observed limits by CMS~\cite{CMS:2017pet}, we estimate the mass bound on $M_{H^{\pm\pm}}$ to be at most 300 GeV for the channels with tau lepton, given the benchmark decay branching ratios in Table~\ref{BR}.

\begin{figure}[h!]
\begin{center}
\includegraphics[scale=1,width=8cm]{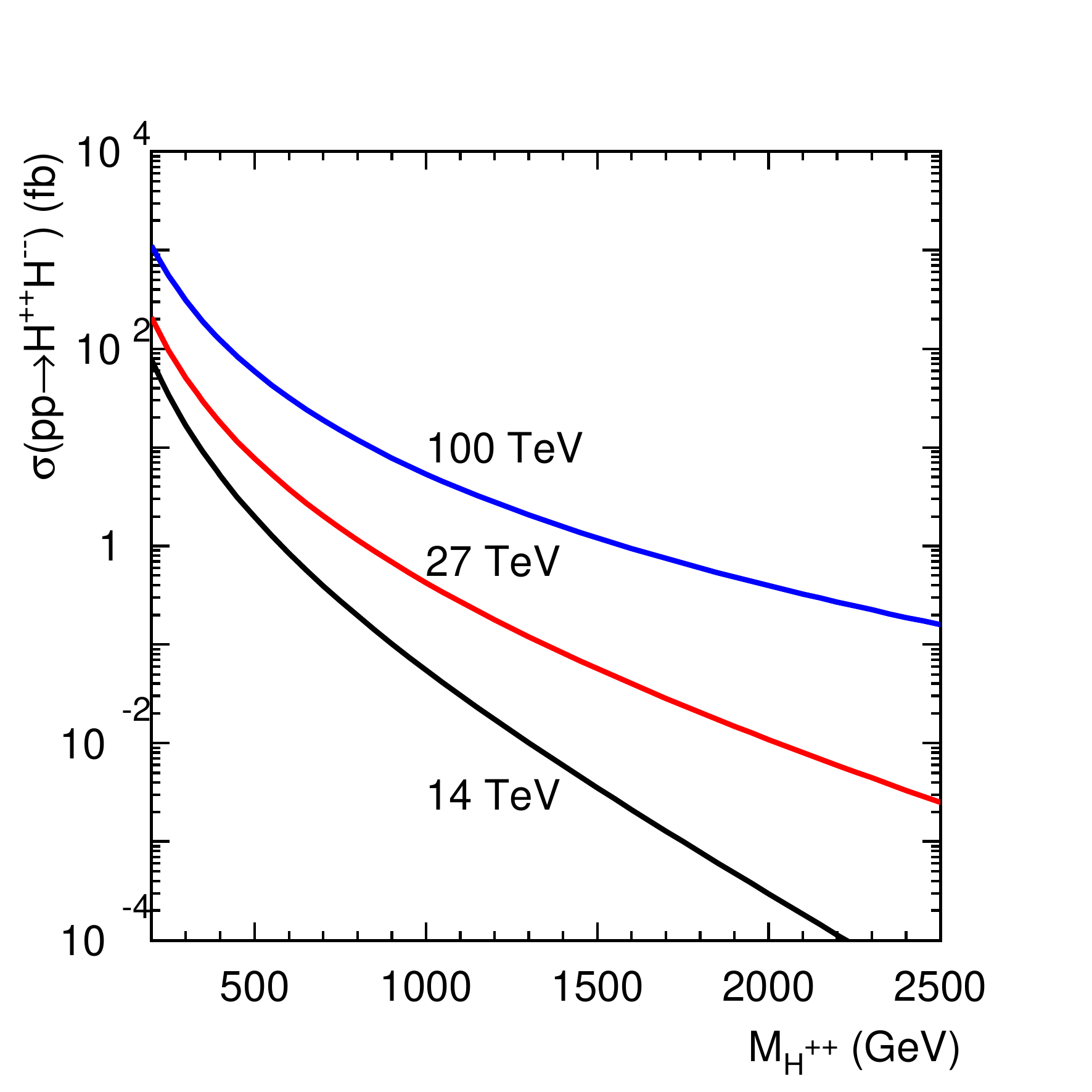}
\end{center}
\caption{The total cross section of $pp\to H^{++} H^{--}$ with collision energy of 14 TeV, 27 TeV and 100 TeV.}
\label{typeii-xsec}
\end{figure}

\subsubsection{$H^{++}H^{--}\to \tau^\pm\ell^\pm\ell^\mp\ell^\mp$}

The first signal channel we consider is the pair production of doubly charged Higgs with one tau lepton in the subsequent decays: $pp\to H^{++}H^{--}\to \tau^\pm\ell^\pm\ell^\mp\ell^\mp$.
Note that we adopt the $\tau$'s leading 2-body decay channel, i.e. $\tau^\pm\to \pi^\pm \overset{(-)}{\nu_\tau}$, with the branching fraction being ${\rm BR}(\tau^\pm\to \pi^\pm \overset{(-)}{\nu_\tau})=0.11$. The $\tau$-spin correlation is maximized in this decay channel. The major SM backgrounds are thus from
\begin{itemize}
\item $ZZ$ diboson events with
one $Z$ decaying to charged leptons $\ell=e, \mu$ and the other to tau lepton pairs.
One tau in the pairs has hadronic 2-body decay and the other tau decays via 3-body leptonic decay mode, i.e. $\tau^\pm\to \ell^\pm\nu_\ell \nu_\tau$:
\begin{eqnarray}
ZZ\to \ell^+\ell^- \tau^+\tau^-\to \ell^+\ell^- \ell^\mp \pi^\pm + \nu's. 
\end{eqnarray}
\item $t\bar{t}Z$ events with $Z$ boson decaying to charged leptons $\ell=e, \mu$. The $W$ boson from one top quark leptonically decays to $\ell^\pm \nu$ and the other $W$ is followed by decay to tau lepton and tau's hadronic decay:
\begin{eqnarray}
t\bar{t}Z_{\to \ell^+\ell^-}\to b\bar{b}\tau^\pm \ell^\mp \ell^+\ell^- + 2\nu\to b\bar{b}\pi^\pm \ell^\mp \ell^+\ell^- + \nu's. 
\end{eqnarray}
\item $t\bar{t}Z$ events with $Z\to \tau^+\tau^-$ and the $W$ bosons' leptonic decay from two top quarks.
The subsequent channels of the tau leptons are respectively through the 2-body hadronic decay and the 3-body leptonic decay as above:
\begin{eqnarray}
t\bar{t}Z_{\to \tau^+\tau^-}\to b\bar{b}\ell^+\ell^-\tau^+\tau^- + 2\nu\to b\bar{b}\ell^+\ell^-\pi^\pm \ell^\mp + \nu's. 
\end{eqnarray}
\end{itemize}
We select events wtih three of the first two generation charged leptons $\ell=e,\mu$ and one hadronically decaying $\tau$ lepton. The three $\ell$ leptons are composed of two same-sign and one opposite-sign leptons. They and the decay product $\pi^\pm$ from tau should satisfy the following basic cuts~\cite{Aaboud:2017qph,CMS:2017pet}:
\begin{equation}
p_T(\ell)\geq 15 \ {\rm GeV}, \ |\eta(\ell)|<2.5; \ p_T(\pi)\geq 20 \ {\rm GeV}, \ |\eta(\pi)|<2.3;\ \Delta R_{\ell \pi}, \Delta R_{\ell\ell} \geq 0.4.
\label{basic1-tau}
\end{equation}

\begin{figure}[h!]
\begin{center}
\minigraph{7.5cm}{-0.15in}{(a)}{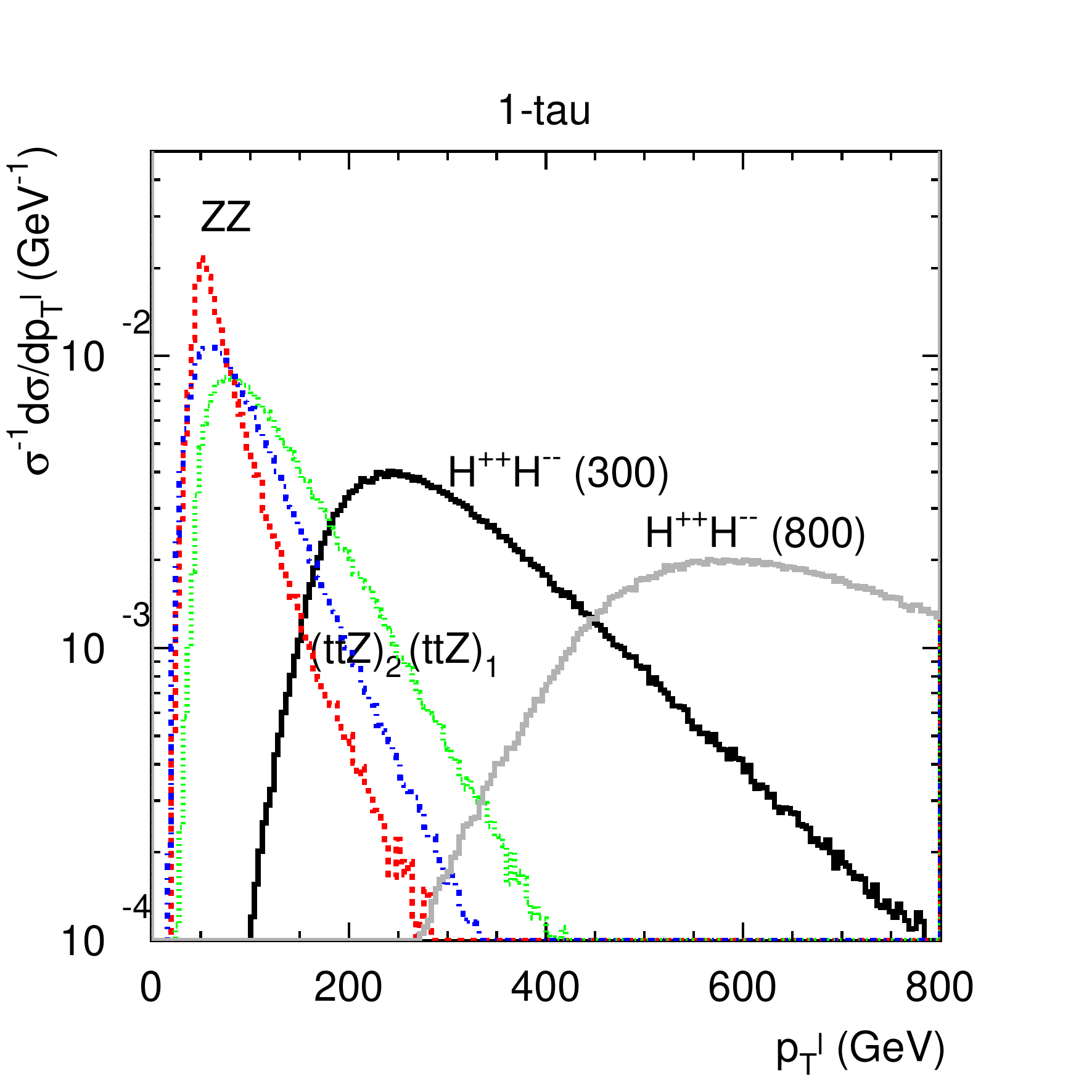}
\minigraph{7.5cm}{-0.15in}{(b)}{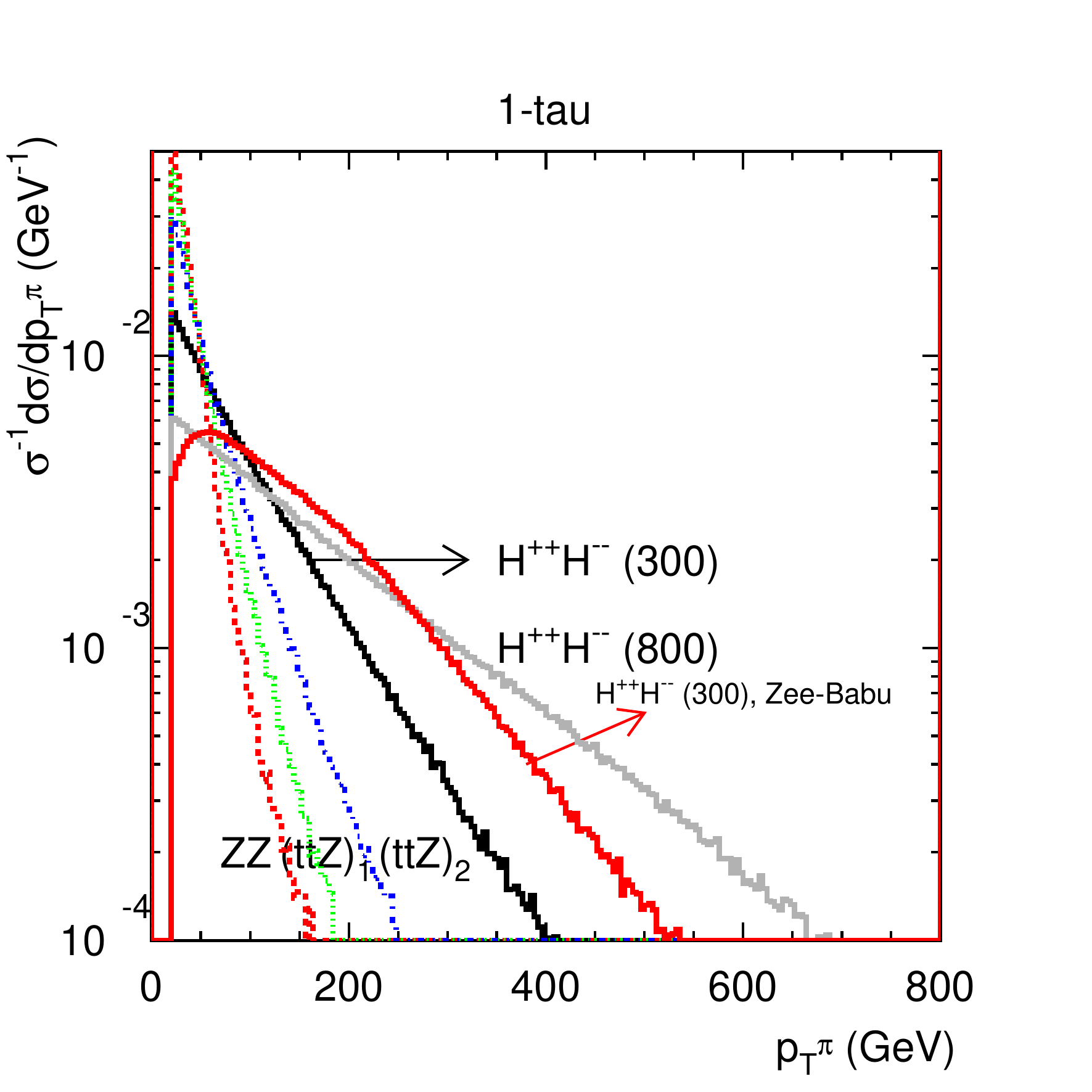}
\minigraph{7.5cm}{-0.15in}{(c)}{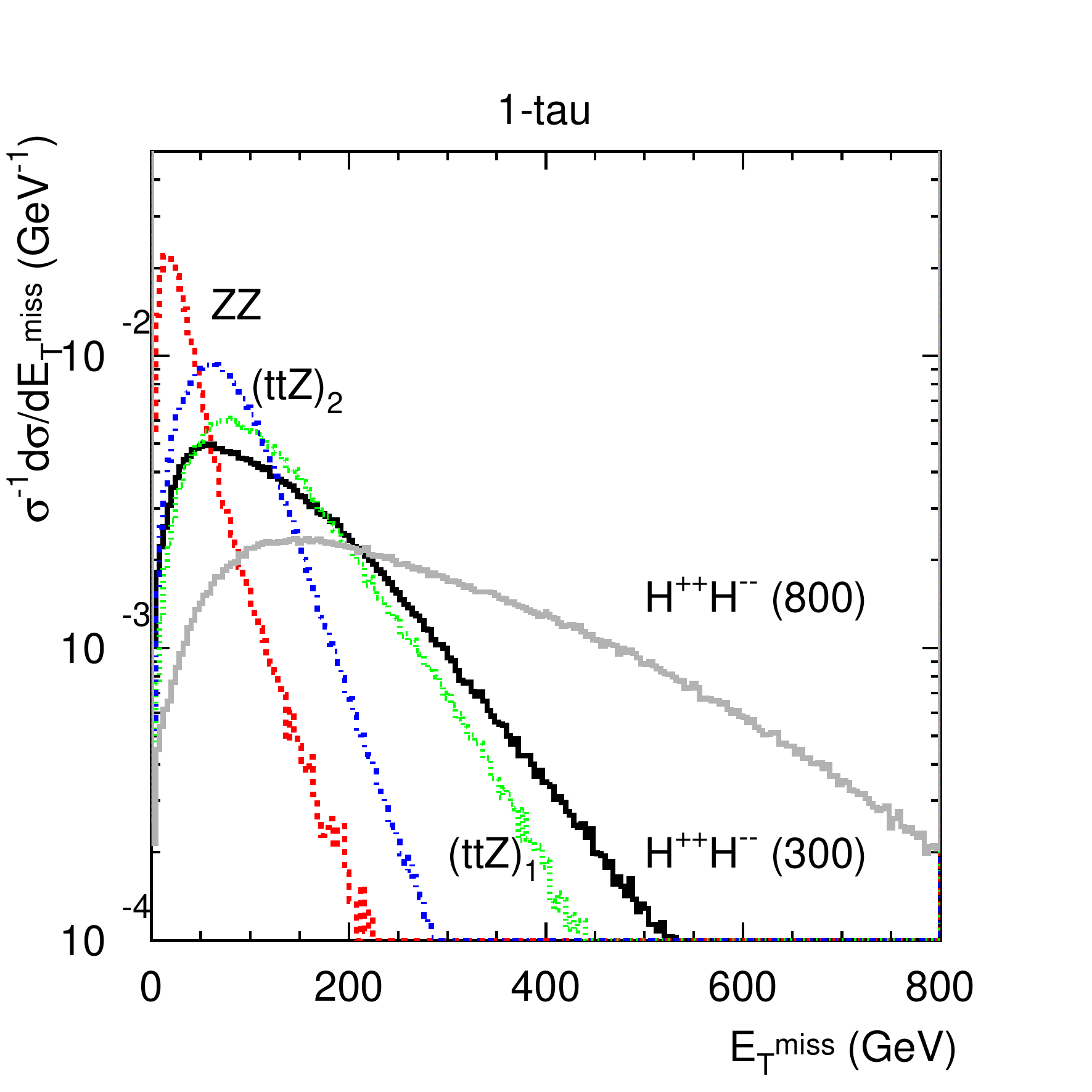}
\minigraph{7.5cm}{-0.15in}{(d)}{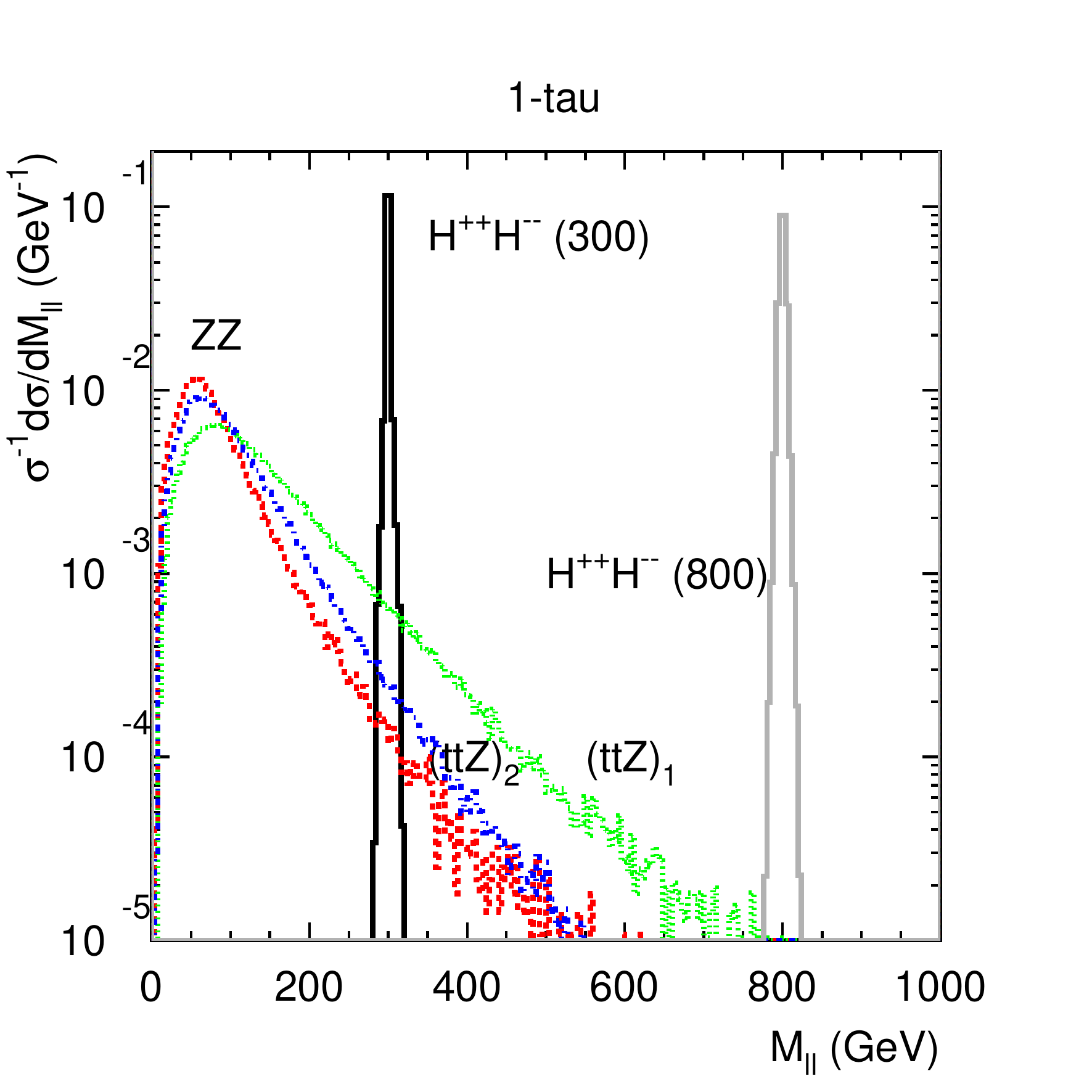}
\end{center}
\caption{The normalized differential cross section distributions of signal $pp\to H^{++}H^{--}\to \tau^\pm\ell^\pm\ell^\mp\ell^\mp$ with $\tau^\pm\to \pi^\pm \overset{(-)}{\nu_\tau}$ and backgrounds versus (a) $p_T^{\rm max}(\ell)$, (b) $p_T(\pi)$, (c) $\cancel{E}_T$ and (d) $M_{\ell^\pm\ell^\pm}$ at the 14 TeV LHC. We assume $M_{H^{\pm\pm}}=300$ and 800 GeV.}
\label{1taudis}
\end{figure}

In Fig.~\ref{1taudis} we display the distributions of signal (assuming $M_{H^{\pm\pm}}=300$ and 800 GeV) and backgrounds at the
14 TeV LHC after the basic cuts shown in Eq.~(\ref{basic1-tau}), for (a) transverse momentum for the hardest $\ell$ lepton,
(b) transverse pion momentum $p_T(\pi)$ and (c) missing transverse energy $\cancel{E}_T$.
As the three $\ell$ leptons are from the decay of heavy doubly charged Higgs in our signal, we tighten the selection cuts by imposing:
\begin{eqnarray}
p_T^{\rm max}(\ell)>M_{H^{\pm\pm}}/2.
\end{eqnarray}
This is from a smeared-out distribution of our signal around the Jacobean peak at $p_T(\ell)\sim M_{\ell^\pm\ell^\pm}/2$.
Furthermore, the signal also has a harder $p_T(\pi)$ spectrum compared to the background. We thus strengthen their transverse momenta and the missing transverse energy induced by tau decay:
\begin{eqnarray}
p_T(\pi), \ \cancel{E}_T>M_{H^{\pm\pm}}/30+40 \ {\rm GeV}.
\end{eqnarray}
Note that, for the $H^{--}$ from an SU$(2)_L$ singlet field coupled only with right-handed tau lepton $\tau^-_R$, the right-handed $\tau$ decays to a left-handed $\nu_\tau$, causing the $\pi^-$ to preferentially move along the $\tau^-$ momentum direction. In contrast, the
$\tau^-$ coming from a triplet $H^{--}$ decay is left-handed which has the opposite effect on the $\pi^-$. The similar feature holds for the $\tau^+$ from $H^{++}$ decay. This is a well-known result of spin
correlation in the $\tau$ decay~\cite{Bullock:1991fd,Bullock:1992yt}. Thus, the transverse momentum of $\pi^\pm$ from doubly charged Higgs decay to tau lepton in Zee-Babu model yields a harder spectrum than that from Type II Seesaw, as seen in Fig.~\ref{1taudis} (b).

Both the $ZZ$ and the 1st $t\bar{t}Z$ backgrounds have the opposite-sign lepton pairs $\ell^+\ell^-$ from $Z$ boson decay and can thus be reduced by
vetoing the invariant mass of opposite-sign leptons satisfying
\begin{eqnarray}
|M_{\ell^+\ell^-}-M_Z|<10 \ {\rm GeV}.
\end{eqnarray}
The invariant mass of the same-sign leptons after basic cuts is shown in Fig.~\ref{1taudis} (d). One can further apply a doubly charged Higgs resonance window, i.e. $|M_{\ell^\pm\ell^\pm}-M_{H^{\pm\pm}}|<0.1M_{H^{\pm\pm}}$, to discover signal events. The $t\bar{t}Z$ backgrounds can also be reduced by vetoing extra b-jets and we find the backgrounds are at negligible level after these requirements.


\begin{table}[tb]
\begin{center}
\begin{tabular}{|c|c|c|c|c|c|}
\hline
$1\tau$ efficiencies  & basic cuts & $p_T^{\ell}$ & $p_T^\pi+\cancel{E}_T$ & $Z$ veto & $M_{\ell^\pm\ell^\pm}$
\\ \hline
$H^{++} H^{--} (300)$ & 0.557 & 0.543 & 0.3 & 0.2834  & 0.2834   \\
$H^{++} H^{--} (800)$ & 0.762 & 0.726 & 0.503 & 0.5021  & 0.5021   \\
\hline \hline
$ZZ$ (300) & 0.07 & 0.00456 & 0.00227 & $1.27\times 10^{-4}$ & $4\times 10^{-6}$   \\
$ZZ$ (800) & 0.07 & $4.8\times 10^{-5}$ & $2.7\times 10^{-5}$ & $<1\times 10^{-6}$ & $<1\times 10^{-6}$   \\
\hline
$(t\bar{t}Z)_1$ (300) & 0.195 & 0.056 & 0.019  & $9.6\times 10^{-4}$  & $1\times 10^{-4}$   \\
$(t\bar{t}Z)_1$ (800) & 0.195 & 0.0017 & $5.38\times 10^{-4}$  & $2.9\times 10^{-5}$  & $1\times 10^{-6}$   \\
\hline
$(t\bar{t}Z)_2$ (300) & 0.1685 & 0.0235 & 0.0097  & 0.00846  & $5.2\times 10^{-4}$ \\
$(t\bar{t}Z)_2$ (800) & 0.1685 & $4.1\times 10^{-4}$ & $1.6\times 10^{-4}$  & $1.5\times 10^{-4}$  & $<1\times 10^{-6}$ \\
\hline  
\end{tabular}
\end{center}
\caption{The cut efficiencies for $1\tau$ signal ($pp\to H^{++}H^{--}\to \tau^\pm\ell^\pm\ell^\mp\ell^\mp$) and the SM backgrounds after accumulated cuts with $\tau^\pm\to \pi^\pm \overset{(-)}{\nu_\tau}$ channel at the 14 TeV LHC. We assume $M_{H^{\pm\pm}}=300$ or 800 GeV.}
\label{1tau}
\end{table}

\begin{figure}[h!]
\begin{center}
\minigraph{7.5cm}{-0.05in}{(a)}{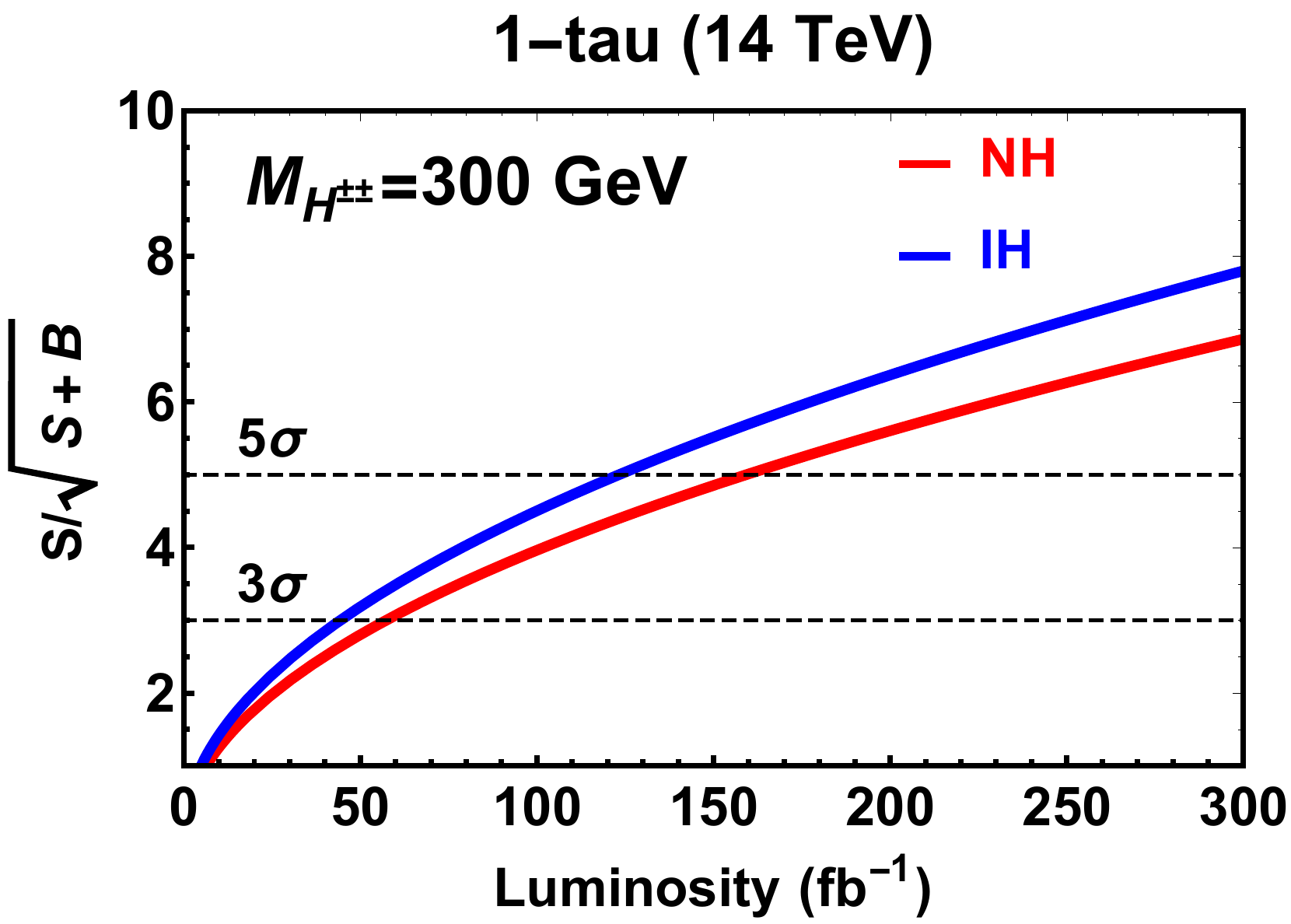}
\minigraph{7.5cm}{-0.05in}{(b)}{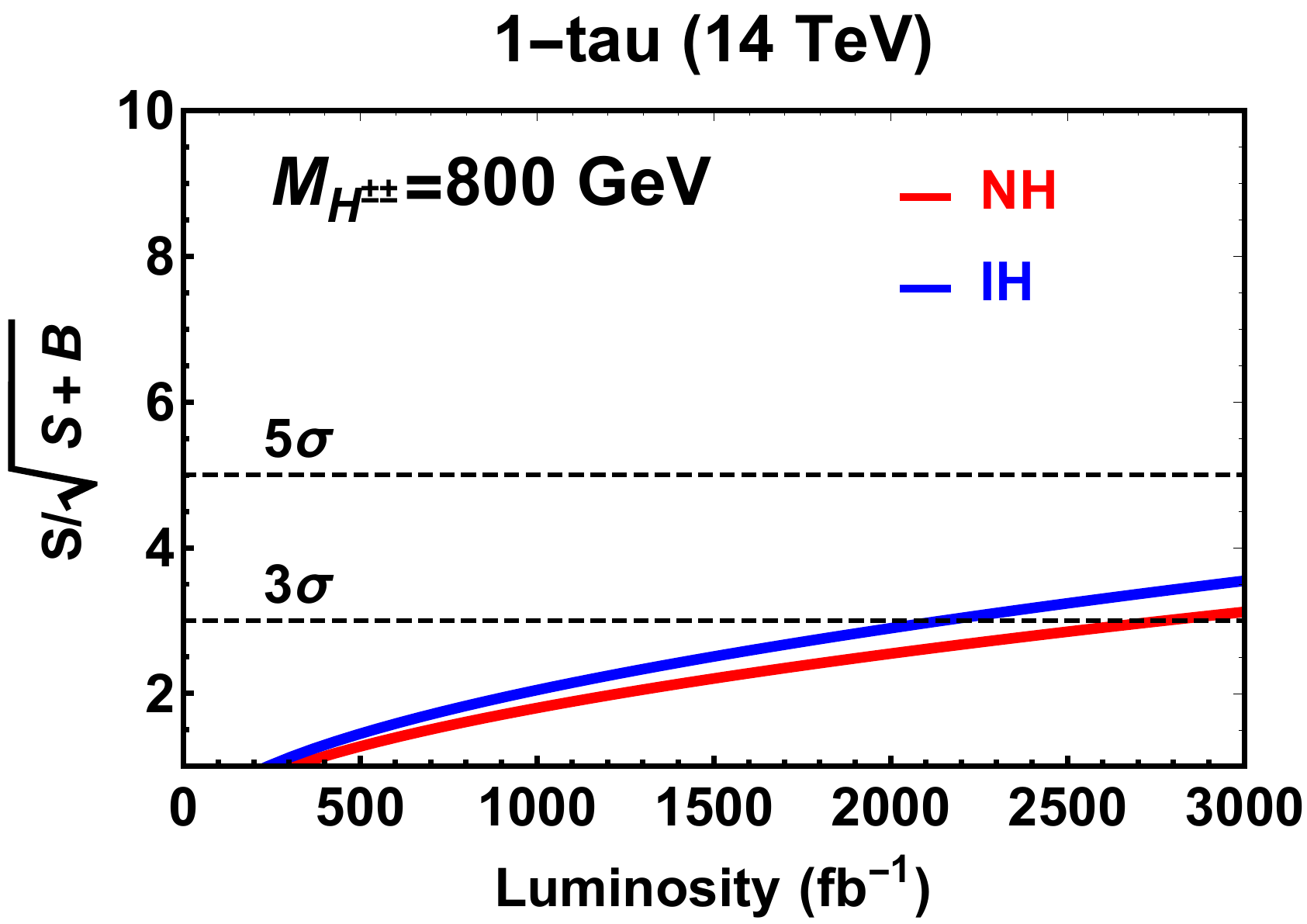}
\end{center}
\caption{Significance of $pp\to H^{++}H^{--}\to \tau^\pm\ell^\pm\ell^\mp\ell^\mp$ with $\tau^\pm\to \pi^\pm \overset{(-)}{\nu_\tau}$ versus integrated luminosity ($\rm fb^{-1}$) for NH and IH at 14 TeV LHC. We assume $M_{H^{\pm\pm}}=$ 300 and 800 GeV.}
\label{sig1tau}
\end{figure}

Given the characteristic features of the signal and backgrounds discussed above, we can use the Gaussian method to calculate the significance
\begin{eqnarray}
S/\sqrt{S+B}
\end{eqnarray}
where $S$ is the signal expectation and $B$ is the sum of background expectations $B=\sum_{i=1,2,3} B_i$. With the proper branching fractions the discovered signal events for this channel read as
\begin{eqnarray}
&&S=L\times \sigma(pp\to H^{++}H^{--})\times {\rm BR}(H^{++}\to \tau^+\ell^+)\times {\rm BR}(H^{--}\to \ell^-\ell^-)\times 2\times\nonumber \\
&&{\rm BR}(\tau^+\to \pi^+\bar{\nu}_\tau)\times \epsilon^{1\tau}_{H^{\pm\pm}},
\end{eqnarray}
and the background expectations are
\begin{eqnarray}
&&B_1=L\times \sigma(pp\to ZZ)\times {\rm BR}(Z\to \ell^+\ell^-)\times {\rm BR}(Z\to \tau^+\tau^-)\times 2\times\nonumber \\
&&{\rm BR}(\tau^+\to \pi^+\bar{\nu}_\tau)\times {\rm BR}(\tau^-\to \ell^-\nu_\tau\bar{\nu}_\ell)\times \epsilon^{1\tau}_{ZZ}, \nonumber \\
&&B_2=L\times \sigma(pp\to t\bar{t}Z)\times {\rm BR}(Z\to \ell^+\ell^-)\times \nonumber \\
&&{\rm BR}(W^+\to \ell^+\nu_\ell)\times {\rm BR}(W^-\to \tau^-\bar{\nu}_\tau)\times 2\times {\rm BR}(\tau^-\to \pi^-\nu_\tau)\times \epsilon^{1\tau}_{(t\bar{t}Z)_1}, \nonumber \\
&&B_3=L\times \sigma(pp\to t\bar{t}Z)\times {\rm BR}(Z\to \tau^+\tau^-)\times {\rm BR}(W^+\to \ell^+\nu_\ell)\times {\rm BR}(W^-\to \ell^-\bar{\nu}_\ell)\times \nonumber \\
&&2\times {\rm BR}(\tau^+\to \pi^+\bar{\nu}_\tau)\times {\rm BR}(\tau^-\to \ell^-\nu_\tau\bar{\nu}_\ell)\times \epsilon^{1\tau}_{(t\bar{t}Z)_2},
\end{eqnarray}
where $L$ is the integrated luminosity and the factor of 2 accounts for the charge-conjugation of final states. The $\epsilon^{1\tau}_{H^{\pm\pm}}, \epsilon^{1\tau}_{ZZ}, \epsilon^{1\tau}_{(t\bar{t}Z)_1}, \epsilon^{1\tau}_{(t\bar{t}Z)_2}$ denote the selection efficiencies for our signal with one tau and the relevant backgrounds, respectively, read from Table~\ref{1tau}. Figure~\ref{sig1tau} shows the significance versus integrated luminosity for NH and IH at 14 TeV LHC. We need 60 (45) fb$^{-1}$ and 160 (125) fb$^{-1}$ luminosity to reach $3\sigma$ and $5\sigma$ significance for $M_{H^{\pm\pm}}=300$ GeV in the case of NH (IH) respectively. For $M_{H^{\pm\pm}}=800$ GeV, the luminosity of 2800 (2150) fb$^{-1}$ is required to reach $3\sigma$ significance for NH (IH).

\subsubsection{$H^{++}H^{--}\to \ell^+\tau^+\ell^-\tau^-$}

For the production of $H^{++}H^{--}\to \ell^+\tau^+\ell^-\tau^-$, we require that both tau leptons hadronically decay to charged pion and neutrino.
The signal is thus composed of a pair of opposite-sign charged leptons and pions plus neutrinos, and the SM backgrounds are
\begin{eqnarray}
ZZ\to \ell^+\ell^-\tau^+\tau^-, \ t\bar{t}Z_{\to \ell^+\ell^-}\to b\bar{b}\tau^+\tau^-\nu_\tau \bar{\nu}_\tau \ell^+\ell^-, \ t\bar{t}Z_{\to \tau^+\tau^-}\to b\bar{b}\ell^+\ell^-\nu_\ell \bar{\nu}_\ell \tau^+\tau^-
\end{eqnarray}
followed by $\tau^\pm\to \pi^\pm \overset{(-)}{\nu_\tau}$. We apply the same basic cuts as Eq.~(\ref{basic1-tau}) to the two pairs of opposite-sign $\ell$ lepton and charged pion in this channel.

\begin{figure}[h!]
\begin{center}
\minigraph{7.5cm}{-0.15in}{(a)}{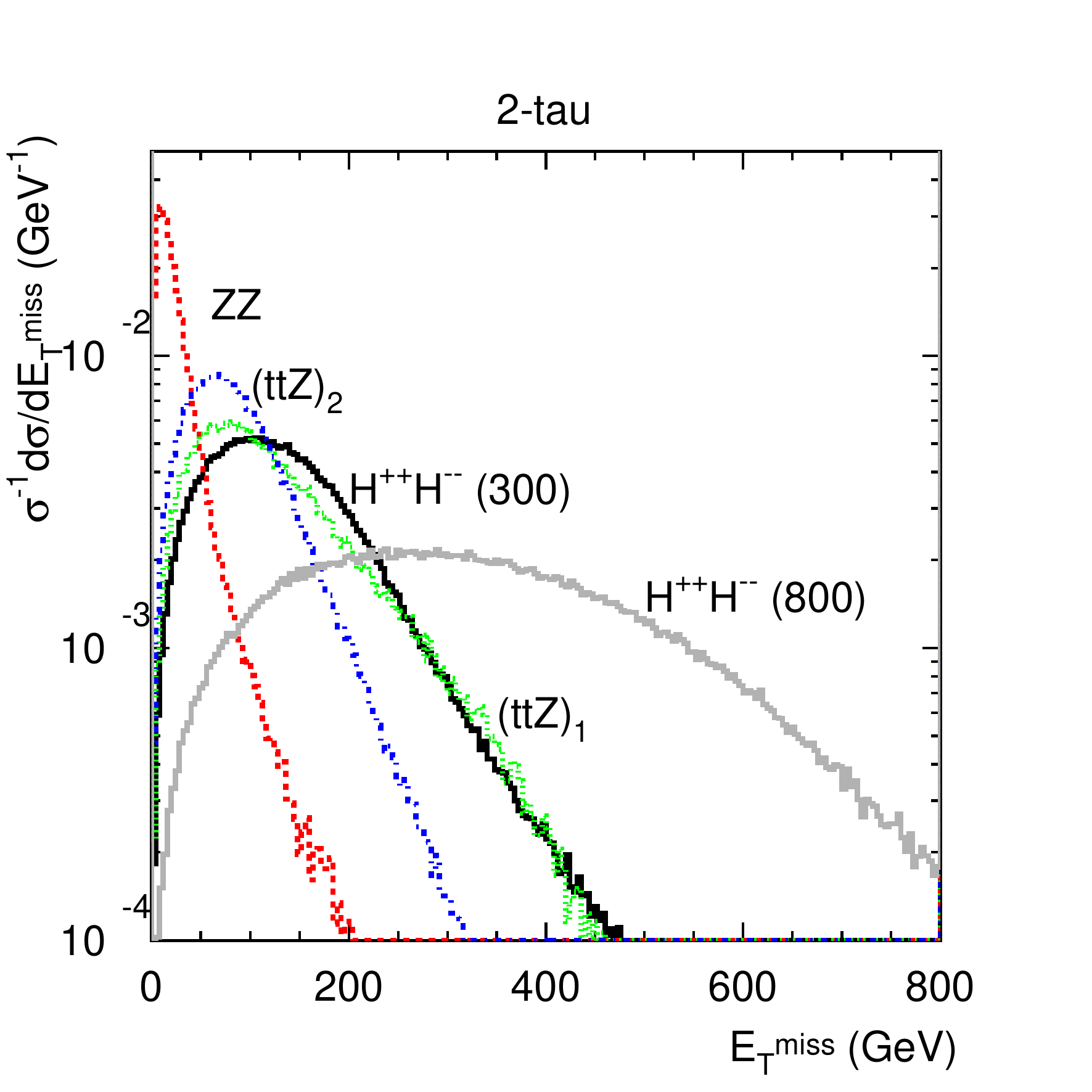}
\minigraph{7.5cm}{-0.15in}{(b)}{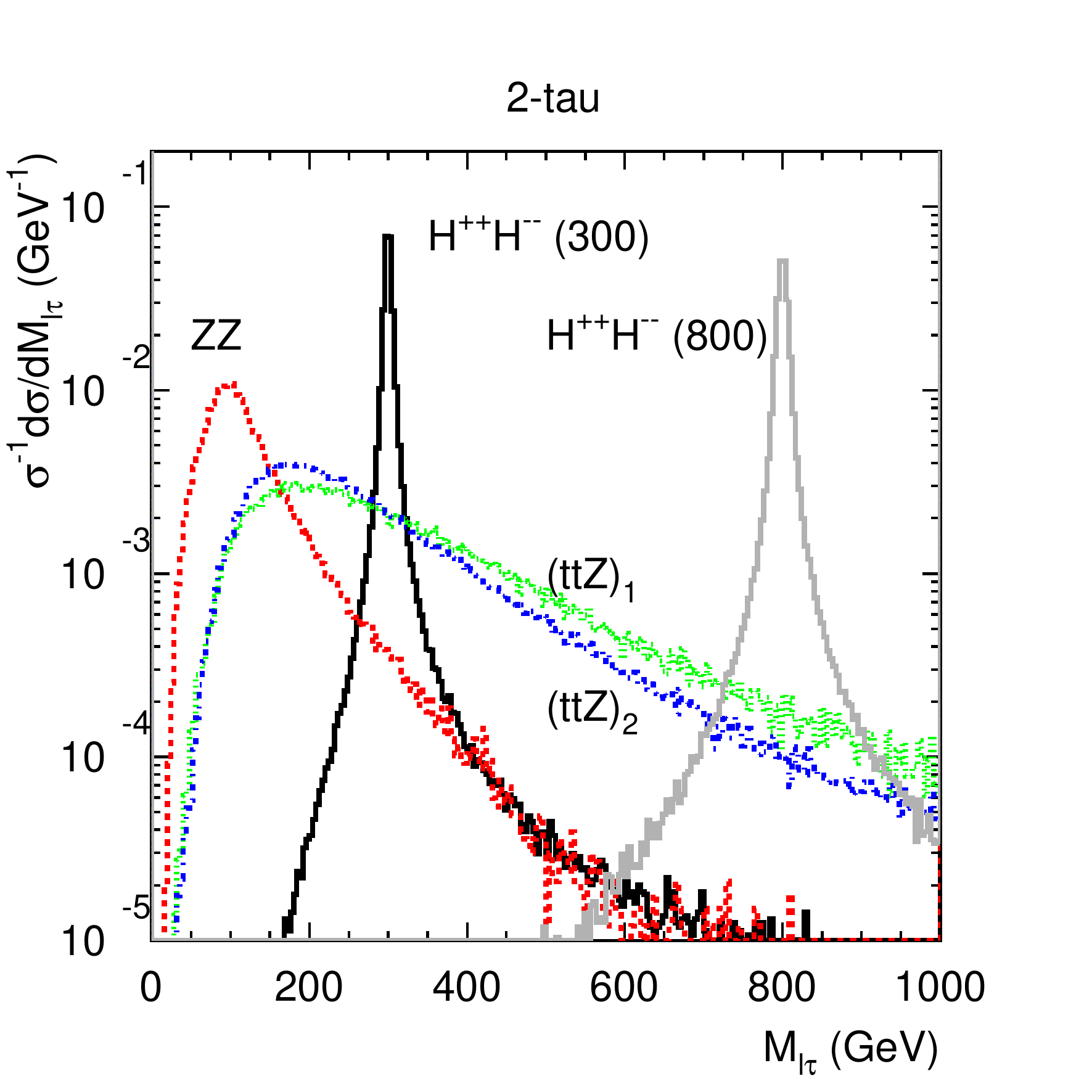}
\end{center}
\caption{The normalized differential cross section distributions of signal $pp\to H^{++}H^{--}\to \ell^+\tau^+\ell^-\tau^-$ with $\tau^\pm\to \pi^\pm \overset{(-)}{\nu_\tau}$ and backgrounds versus (a) $\cancel{E}_T$ and (b) $M_{\ell^\pm\tau^\pm}$ at the 14 TeV LHC. We assume $M_{H^{\pm\pm}}=300$ and 800 GeV.}
\label{2taudis}
\end{figure}

As seen from Fig.~\ref{2taudis} (a), the missing energy is harder in this channel as there are more invisible neutrinos from tau decay. That is also why the kinematical reconstruction of the same-sign dilepton invariant mass is more complicated. Fortunately all the tau leptons are very energetic from the decay of a heavy doubly charged Higgs boson, the missing momentum of neutrinos will be along the direction with the charged track. We thus have
\begin{eqnarray}
\vec{p}({\rm invisible})=k_1 \ \vec{p}({\rm track}_1)+k_2 \ \vec{p}({\rm track}_2)
\end{eqnarray}
where $k_1$ and $k_2$ can be determined from the $\cancel{p}_T$ measurement. The tau and $H^{\pm\pm}$ Higgs pairs' kinematics can thus be fully reconstructed. The reconstructed Higgs mass from $\ell^\pm\tau^\pm$ is shown in Fig.~\ref{2taudis} (b). One can see that the invariant mass is broader than that in the $1\tau$ events. We thus modify the $\cancel{E}_T$ and mass window cut as below
\begin{eqnarray}
\cancel{E}_T>M_{H^{\pm\pm}}/30+60 \ {\rm GeV} \ \ {\rm and} \ \ |M_{\ell^\pm\tau^\pm}-M_{H^{\pm\pm}}|<0.2M_{H^{\pm\pm}}.
\end{eqnarray}
The rest selection requirements are the same as those in $1\tau$ case.

\begin{table}[tb]
\begin{center}
\begin{tabular}{|c|c|c|c|c|c|}
\hline
$2\tau$ efficiencies  & basic cuts & $p_T^{\ell}$ & $p_T^\pi+\cancel{E}_T$ & $Z$ veto & $M_{\ell^\pm\tau^\pm}$
\\ \hline
$H^{++} H^{--} (300)$ & 0.41 & 0.33 & 0.2367 & 0.2321  &  0.228 \\
$H^{++} H^{--} (800)$ & 0.67 & 0.51 & 0.454 & 0.454  & 0.45  \\
\hline \hline
$ZZ$ (300) & 0.0838 & 0.00474 & 0.0025 & $1.5\times 10^{-4}$ & $6\times 10^{-5}$   \\
$ZZ$ (800) & 0.0838 & $4.2\times 10^{-5}$ & $3.4\times 10^{-5}$ & $3\times 10^{-6}$ & $<1\times 10^{-6}$   \\
\hline
$(t\bar{t}Z)_1$ (300) & 0.0733 & 0.019 & 0.01 & $6.27\times 10^{-4}$  & $1.69\times 10^{-4}$   \\
$(t\bar{t}Z)_1$ (800) & 0.0733 & $6.93\times 10^{-4}$ & $3.38\times 10^{-4}$ & $2\times 10^{-5}$  & $9\times 10^{-6}$   \\
\hline
$(t\bar{t}Z)_2$ (300) & 0.1847 & 0.0231 & 0.0127 & 0.0123  & 0.00336 \\
$(t\bar{t}Z)_2$ (800) & 0.1847 & $4\times 10^{-4}$ & $2.52\times 10^{-4}$ & $2.51\times 10^{-4}$  & $7.57\times 10^{-5}$ \\
\hline
\end{tabular}
\end{center}
\caption{The cut efficiencies for $2\tau$ signal ($pp\to H^{++}H^{--}\to \ell^+\tau^+\ell^-\tau^-$) and the SM backgrounds after accumulated cuts with $\tau^\pm\to \pi^\pm \overset{(-)}{\nu_\tau}$ channel at the 14 TeV LHC. We assume $M_{H^{\pm\pm}}=300$ or 800 GeV.}
\label{2tau}
\end{table}

\begin{figure}[h!]
\begin{center}
\includegraphics[scale=1,width=7.5cm]{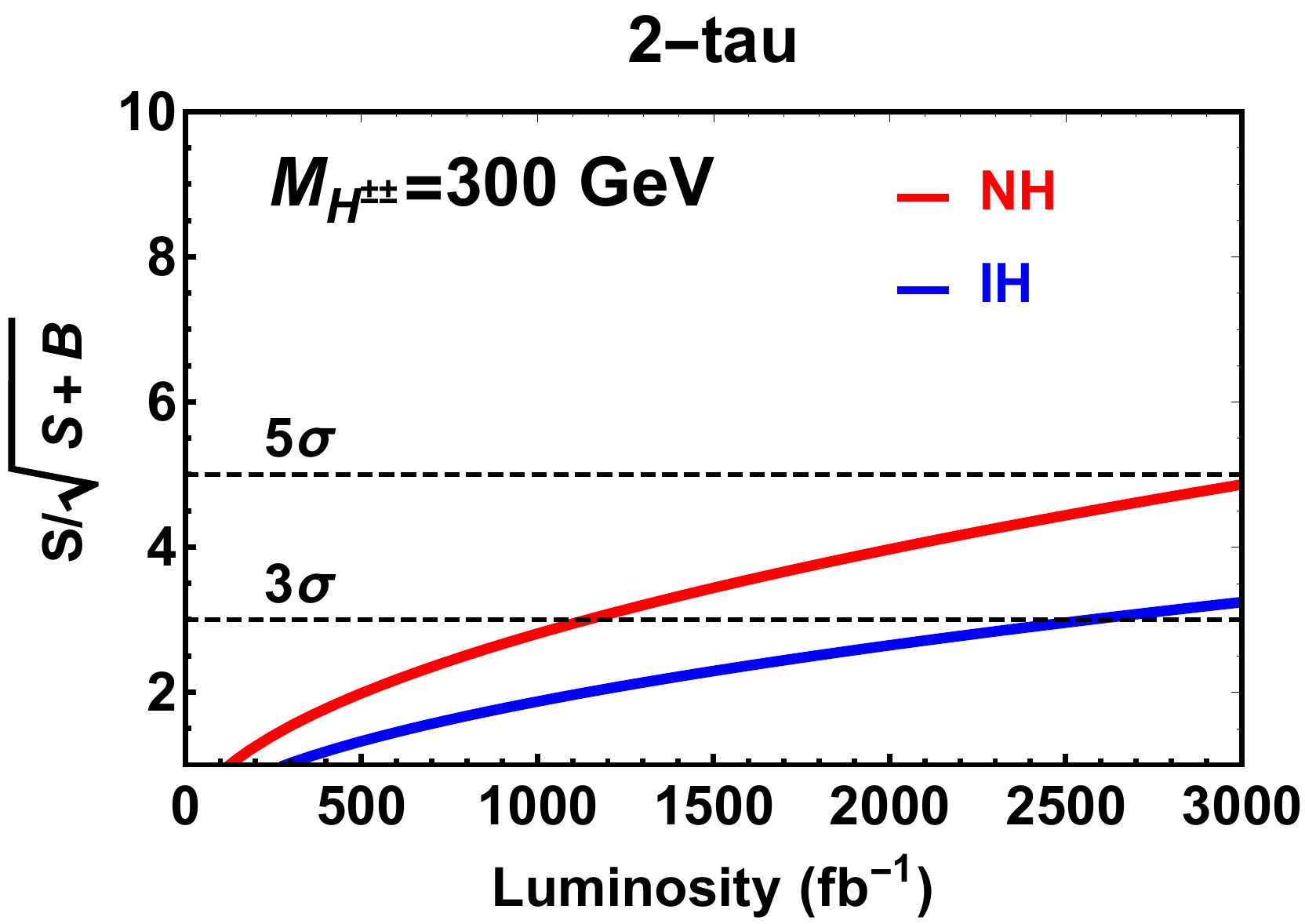}
\end{center}
\caption{Significance of $pp\to H^{++}H^{--}\to \ell^+\tau^+\ell^-\tau^-$ with $\tau^\pm\to \pi^\pm \overset{(-)}{\nu_\tau}$ versus integrated luminosity ($\rm fb^{-1}$) for NH and IH at 14 TeV LHC. We assume $M_{H^{\pm\pm}}=$ 300 GeV.}
\label{sig2tau}
\end{figure}

The discovered signal events for this $2\tau$ channel are
\begin{eqnarray}
&&S=L\times \sigma(pp\to H^{++}H^{--})\times {\rm BR}^2(H^{\pm\pm}\to \tau^\pm\ell^\pm)\times {\rm BR}^2(\tau^+\to \pi^+ \bar{\nu}_\tau)\times \epsilon^{2\tau}_{H^{\pm\pm}}
\nonumber \\
\end{eqnarray}
and those for backgrounds are
\begin{eqnarray}
&&B_1=L\times \sigma(pp\to ZZ)\times {\rm BR}(Z\to \ell^+\ell^-)\times {\rm BR}(Z\to \tau^+\tau^-)\times \nonumber \\
&&{\rm BR}^2(\tau^+\to \pi^+ \bar{\nu}_\tau)\times \epsilon^{2\tau}_{ZZ}, \nonumber \\
&&B_2=L\times \sigma(pp\to t\bar{t}Z)\times {\rm BR}(Z\to \ell^+\ell^-)\times {\rm BR}^2(W^\pm\to \tau^\pm \overset{(-)}{\nu_\tau})\times \nonumber \\
&&{\rm BR}^2(\tau^+\to \pi^+ \bar{\nu}_\tau)\times \epsilon^{2\tau}_{(t\bar{t}Z)_1}, \nonumber \\
&&B_3=L\times \sigma(pp\to t\bar{t}Z)\times {\rm BR}(Z\to \tau^+\tau^-)\times {\rm BR}^2(W^\pm\to \ell^\pm \overset{(-)}{\nu_\tau})\times \nonumber \\
&&{\rm BR}^2(\tau^+\to \pi^+ \bar{\nu}_\tau)\times \epsilon^{2\tau}_{(t\bar{t}Z)_2},
\end{eqnarray}
where the $\epsilon^{2\tau}_{H^{\pm\pm}}, \epsilon^{2\tau}_{ZZ}, \epsilon^{2\tau}_{(t\bar{t}Z)_1}, \epsilon^{2\tau}_{(t\bar{t}Z)_2}$ are the cut efficiencies from Table~\ref{2tau}. From Fig.~\ref{sig2tau}, one can see that the luminosity of 1150 (2600) fb$^{-1}$ is required to reach $3\sigma$ significance for NH (IH) when $M_{H^{\pm\pm}}=300$ GeV. For this channel, it is impossible to discover heavier doubly charged Higgs at 14 TeV LHC with luminosity below 3000 fb$^{-1}$.
Upgrades of the LHC are needed to probe the doubly charged Higgs with decay to tau lepton.

\section{Results for the HL-LHC, HE-LHC and FCC-hh}
\label{sec:results}

In this section we show the results of projected sensitivity to leptonic processes with tau lepton in Type II Seesaw, at the high luminosity/energy upgrades of the LHC and the future FCC-hh.
For the HE-LHC and FCC-hh, we require that the decay products of doubly charged Higgs satisfy the following basic cuts:
\begin{equation}
p_T(\ell)\geq 30 \ {\rm GeV}, \ p_T(\pi)\geq 25 \ {\rm GeV}; \ |\eta(\ell,\pi)|<2.5;\ \Delta R_{\ell \pi}, \Delta R_{\ell\ell} \geq 0.4.
\label{basic2-tau}
\end{equation}
The rest of selection cuts are the same as above.

\begin{figure}[h!]
\begin{center}
\includegraphics[scale=1,width=8.5cm]{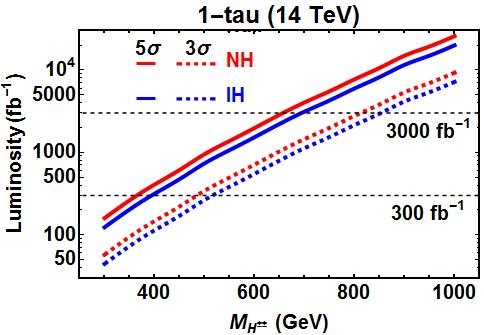}
\end{center}
\caption{Integrated luminosity versus $M_{H^{\pm\pm}}$ at $3\sigma$ and $5\sigma$ significance for $pp\to H^{++}H^{--}\to \tau^\pm\ell^\pm\ell^\mp\ell^\mp$ with $\tau^\pm\to \pi^\pm \overset{(-)}{\nu_\tau}$ for NH and IH at 14 TeV LHC. }
\label{LMtau14}
\end{figure}

\begin{figure}[h!]
\begin{center}
\minigraph{7.5cm}{-0.05in}{(a)}{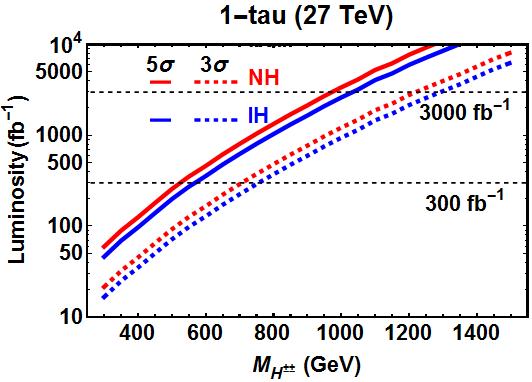}
\minigraph{7.5cm}{-0.05in}{(b)}{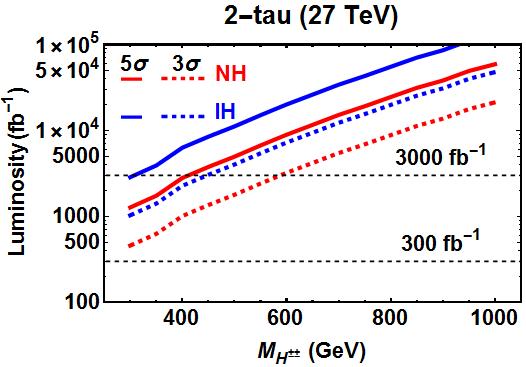}
\end{center}
\caption{Integrated luminosity versus $M_{H^{\pm\pm}}$ at $3\sigma$ and $5\sigma$ significance for $pp\to H^{++}H^{--}\to \tau^\pm\ell^\pm\ell^\mp\ell^\mp$ (a) and $\ell^+\tau^+\ell^-\tau^-$ (b) with $\tau^\pm\to \pi^\pm \overset{(-)}{\nu_\tau}$ for NH and IH at 27 TeV LHC. }
\label{LMtau27}
\end{figure}

\begin{figure}[h!]
\begin{center}
\minigraph{7.5cm}{-0.05in}{(a)}{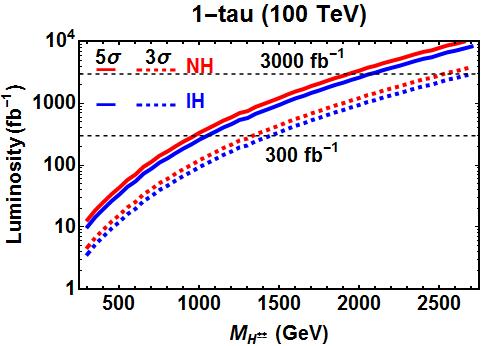}
\minigraph{7.5cm}{-0.05in}{(b)}{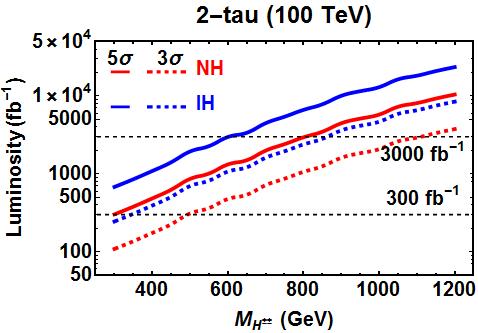}
\end{center}
\caption{Integrated luminosity versus $M_{H^{\pm\pm}}$ at $3\sigma$ and $5\sigma$ significance for $pp\to H^{++}H^{--}\to \tau^\pm\ell^\pm\ell^\mp\ell^\mp$ (a) and $\ell^+\tau^+\ell^-\tau^-$ (b) with $\tau^\pm\to \pi^\pm \overset{(-)}{\nu_\tau}$ for NH and IH at 100 TeV LHC. }
\label{LMtau100}
\end{figure}

In Figs.~\ref{LMtau14}, \ref{LMtau27} and \ref{LMtau100} we display the 3$\sigma$ and 5$\sigma$ significance in the plane of integrated luminosity versus doubly charged Higgs mass for $pp\to H^{++}H^{--}\to \tau^\pm\ell^\pm\ell^\mp\ell^\mp, \ell^+\tau^+\ell^-\tau^-$ at the 14 TeV LHC, 27 TeV LHC and 100 TeV FCC-hh, respectively. One can see that the 14 TeV LHC can probe $H^{++}$ mass up to 365 (655) GeV and 485 (815) GeV with 5$\sigma$ and 3$\sigma$ significance respectively for NH through $1\tau$ channel, assuming $L=300 \ (3000) \ {\rm fb}^{-1}$, while for IH the masses of 395 (695) GeV and 515 (855) GeV can be achieved. Note that, as taking tau polarization effect into account through its hadronic decay product, our prediction would be more conservative than those LHC made. The discovery mass at the HE-LHC and FCC-hh is above 525 (975) GeV and 960 (1930) GeV respectively, for the luminosity of 300 (3000) fb$^{-1}$. The accessible mass for $2\tau$ channel is much lower than $1\tau$ channel, e.g. 300-400 GeV at HE-LHC and 600-800 GeV at FCC-hh.

Finally, we summarize the reachable doubly charged Higgs mass at 5$\sigma$ significance for $1\tau$ and $2\tau$ channels and NH and IH neutrino mass patterns, at the 14 TeV LHC, 27 TeV LHC and 100 TeV FCC-hh, in Table~\ref{sumsig}.

\begin{table}[tb]
\begin{center}
\begin{tabular}{|c|c|c|c|}
\hline
$M_{H^{\pm\pm}}$ & 14 TeV & 27 TeV & 100 TeV
\\ \hline
$1\tau$ NH & 365 (655) & 525 (975) & 960 (1930) \\
$1\tau$ IH & 395 (695) & 567 (1035) & 1050 (2070) \\
\hline \hline
$2\tau$ NH & -- & -- (410) & 305 (805)  \\
$2\tau$ IH & -- & -- (305) & -- (609)  \\
\hline
\end{tabular}
\end{center}
\caption{The reachable doubly charged Higgs mass (in the unit of GeV) at 5$\sigma$ significance for $1\tau$ and $2\tau$ channels and NH and IH neutrino mass patterns, at the 14 TeV LHC, 27 TeV LHC and FCC-hh. We assume $L=300 \ (3000) \ {\rm fb}^{-1}$.}
\label{sumsig}
\end{table}

\section{Conclusions}
\label{sec:Concl}

Neutrino oscillation measurements indicate that tau lepton plays manifest role in distinguishing different neutrino mass patterns. Moreover, tau polarization can help to determine the chiral property of its parent particle and thus discriminate different heavy scalar mediated neutrino mass mechanisms, such as Type II Seesaw and Zee-Babu model. We thus examine the lepton flavor signatures with tau lepton at LHC upgrades, i.e. HL-LHC, HE-LHC and FCC-hh, through leptonic processes from doubly charged Higgs in the Type II Seesaw. We investigate signal channels with both one and two tau leptons in final states from doubly charged Higgs pair production, i.e. $pp\to H^{++}H^{--}\to \tau^\pm\ell^\pm\ell^\mp\ell^\mp, \ell^+\tau^+\ell^-\tau^-$, and leading SM backgrounds with $\tau^\pm\to \pi^\pm \nu$.

With the analysis presented here we have found that, for the $1\tau$ channel, the projected doubly charged Higgs mass at HL-LHC can reach 655 GeV and 695 GeV for NH and IH respectively with the luminosity of 3000 fb$^{-1}$. Higher masses, 975-1930 GeV for NH and 1035-2070 GeV for IH, can be achieved at HE-LHC and FCC-hh.  The accessible mass for $2\tau$ channel is much lower than that for $1\tau$ channel.

\acknowledgments
The National Computational Infrastructure (NCI), the Southern Hemisphere's fastest supercomputer, is gratefully acknowledged.



\begin{thebibliography}{100}


\bibitem{Weinberg}
  S.~Weinberg,
  ``Baryon And Lepton Nonconserving Processes,''
  Phys.\ Rev.\ Lett.\  {\bf 43} (1979) 1566.


\bibitem{Konetschny:1977bn}
  W.~Konetschny and W.~Kummer,
  Phys.\ Lett.\  {\bf 70B}, 433 (1977).
  doi:10.1016/0370-2693(77)90407-5

\bibitem{Cheng:1980qt}
  T.~P.~Cheng and L.~F.~Li,
  Phys.\ Rev.\ D {\bf 22}, 2860 (1980).
  doi:10.1103/PhysRevD.22.2860

\bibitem{Lazarides:1980nt}
  G.~Lazarides, Q.~Shafi and C.~Wetterich,
  Nucl.\ Phys.\ B {\bf 181}, 287 (1981).
  doi:10.1016/0550-3213(81)90354-0

\bibitem{Schechter:1980gr}
  J.~Schechter and J.~W.~F.~Valle,
  Phys.\ Rev.\ D {\bf 22}, 2227 (1980).
  doi:10.1103/PhysRevD.22.2227

\bibitem{Mohapatra:1980yp}
  R.~N.~Mohapatra and G.~Senjanovic,
  Phys.\ Rev.\ D {\bf 23}, 165 (1981).
  doi:10.1103/PhysRevD.23.165





\bibitem{deGouvea:2006gz}
  A.~de Gouvea, J.~Jenkins and N.~Vasudevan,
  Phys.\ Rev.\ D {\bf 75}, 013003 (2007)
  doi:10.1103/PhysRevD.75.013003
  [hep-ph/0608147].

\bibitem{deGouvea:2007hks}
  A.~de Gouvea,
  arXiv:0706.1732 [hep-ph].



\bibitem{LNVreview}
For recent review on TeV seesaw, see e.g.,
  F.~F.~Deppisch, P.~S.~Bhupal Dev and A.~Pilaftsis,
  New J.\ Phys.\  {\bf 17}, no. 7, 075019 (2015)
  doi:10.1088/1367-2630/17/7/075019
  [arXiv:1502.06541 [hep-ph]],
Y.~Cai, T.~Han, T.~Li and R.~Ruiz,
  arXiv:1711.02180 [hep-ph]
  and references therein.




\bibitem{Aaboud:2017qph}
  M.~Aaboud {\it et al.} [ATLAS Collaboration],
  arXiv:1710.09748 [hep-ex].


\bibitem{Perez:2008ha}
  P.~Fileviez Perez, T.~Han, G.~Y.~Huang, T.~Li and K.~Wang,
  Phys.\ Rev.\ D {\bf 78}, 071301 (2008)
  doi:10.1103/PhysRevD.78.071301
  [arXiv:0803.3450 [hep-ph]];
  P.~Fileviez Perez, T.~Han, G.~y.~Huang, T.~Li and K.~Wang,
  Phys.\ Rev.\ D {\bf 78}, 015018 (2008)
  doi:10.1103/PhysRevD.78.015018
  [arXiv:0805.3536 [hep-ph]].





\bibitem{CMS:2017pet}
  CMS Collaboration [CMS Collaboration],
  CMS-PAS-HIG-16-036.



\bibitem{delAguila:2013mia}
  F.~del \'{A}guila and M.~Chala,
  JHEP {\bf 1403}, 027 (2014)
  doi:10.1007/JHEP03(2014)027
  [arXiv:1311.1510 [hep-ph]].



\bibitem{Zee:1985id}
  A.~Zee,
  Nucl.\ Phys.\ B {\bf 264}, 99 (1986).
  doi:10.1016/0550-3213(86)90475-X

\bibitem{Babu:1988ki}
  K.~S.~Babu,
  Phys.\ Lett.\ B {\bf 203}, 132 (1988).
  doi:10.1016/0370-2693(88)91584-5



\bibitem{Sugiyama:2012yw}
  H.~Sugiyama, K.~Tsumura and H.~Yokoya,
  Phys.\ Lett.\ B {\bf 717}, 229 (2012)
  doi:10.1016/j.physletb.2012.09.044
  [arXiv:1207.0179 [hep-ph]].


\bibitem{Abe:2011fz}
  Y.~Abe {\it et al.} [Double Chooz Collaboration],
  Phys.\ Rev.\ Lett.\  {\bf 108}, 131801 (2012)
  doi:10.1103/PhysRevLett.108.131801
  [arXiv:1112.6353 [hep-ex]].

\bibitem{Ahn:2012nd}
  J.~K.~Ahn {\it et al.} [RENO Collaboration],
  Phys.\ Rev.\ Lett.\  {\bf 108}, 191802 (2012)
  doi:10.1103/PhysRevLett.108.191802
  [arXiv:1204.0626 [hep-ex]].

\bibitem{An:2012eh}
  F.~P.~An {\it et al.} [Daya Bay Collaboration],
  Phys.\ Rev.\ Lett.\  {\bf 108}, 171803 (2012)
  doi:10.1103/PhysRevLett.108.171803
  [arXiv:1203.1669 [hep-ex]].


\bibitem{Abe:2017uxa}
  K.~Abe {\it et al.} [T2K Collaboration],
  Phys.\ Rev.\ Lett.\  {\bf 118}, no. 15, 151801 (2017)
  doi:10.1103/PhysRevLett.118.151801
  [arXiv:1701.00432 [hep-ex]].

\bibitem{Adamson:2016tbq}
  P.~Adamson {\it et al.} [NOvA Collaboration],
  Phys.\ Rev.\ Lett.\  {\bf 116}, no. 15, 151806 (2016)
  doi:10.1103/PhysRevLett.116.151806
  [arXiv:1601.05022 [hep-ex]].

\bibitem{Abe:2013hdq}
  K.~Abe {\it et al.} [T2K Collaboration],
  Phys.\ Rev.\ Lett.\  {\bf 112}, 061802 (2014)
  doi:10.1103/PhysRevLett.112.061802
  [arXiv:1311.4750 [hep-ex]].






\bibitem{Jadach:1990mz}
  S.~Jadach, J.~H.~Kuhn and Z.~Was,
  Comput.\ Phys.\ Commun.\  {\bf 64}, 275 (1990).
  doi:10.1016/0010-4655(91)90038-M

\bibitem{Jezabek:1991qp}
  M.~Jezabek, Z.~Was, S.~Jadach and J.~H.~Kuhn,
  Comput.\ Phys.\ Commun.\  {\bf 70}, 69 (1992).
  doi:10.1016/0010-4655(92)90092-D

\bibitem{Jadach:1993hs}
  S.~Jadach, Z.~Was, R.~Decker and J.~H.~Kuhn,
  Comput.\ Phys.\ Commun.\  {\bf 76}, 361 (1993).
  doi:10.1016/0010-4655(93)90061-G



\bibitem{Hagiwara:2012vz}
  K.~Hagiwara, T.~Li, K.~Mawatari and J.~Nakamura,
  Eur.\ Phys.\ J.\ C {\bf 73}, 2489 (2013)
  doi:10.1140/epjc/s10052-013-2489-4
  [arXiv:1212.6247 [hep-ph]].


\bibitem{Esteban:2016qun}
  I.~Esteban, M.~C.~Gonzalez-Garcia, M.~Maltoni, I.~Martinez-Soler and T.~Schwetz,
  JHEP {\bf 1701}, 087 (2017)
  doi:10.1007/JHEP01(2017)087
  [arXiv:1611.01514 [hep-ph]]; NuFIT 3.2 (2018), www.nu-fit.org.

\bibitem{Ade:2015xua}
  P.~A.~R.~Ade {\it et al.} [Planck Collaboration],
  Astron.\ Astrophys.\  {\bf 594}, A13 (2016)
  doi:10.1051/0004-6361/201525830
  [arXiv:1502.01589 [astro-ph.CO]].





\bibitem{Muhlleitner:2003me}
  M.~Muhlleitner and M.~Spira,
  Phys.\ Rev.\ D {\bf 68}, 117701 (2003)
  doi:10.1103/PhysRevD.68.117701
  [hep-ph/0305288].


\bibitem{Alwall:2014hca}
  J.~Alwall {\it et al.},
  JHEP {\bf 1407}, 079 (2014)
  doi:10.1007/JHEP07(2014)079
  [arXiv:1405.0301 [hep-ph]].



\bibitem{Bullock:1991fd}
  B.~K.~Bullock, K.~Hagiwara and A.~D.~Martin,
  Phys.\ Rev.\ Lett.\  {\bf 67}, 3055 (1991).
  doi:10.1103/PhysRevLett.67.3055

\bibitem{Bullock:1992yt}
  B.~K.~Bullock, K.~Hagiwara and A.~D.~Martin,
  Nucl.\ Phys.\ B {\bf 395}, 499 (1993).
  doi:10.1016/0550-3213(93)90045-Q






\end{thebibliography}
\end{document}